\documentclass[12pt]{elsart}
\usepackage{epsfig}
\usepackage{amssymb}
\usepackage{times}
\begin{document}

\newcommand{\re}{\mathop{\mathrm{Re}}}
\newcommand{\im}{\mathop{\mathrm{Im}}}
\newcommand{\D}{\mathop{\mathrm{d}}}
\newcommand{\I}{\mathop{\mathrm{i}}}

\noindent {\Large DESY 03-197}

\noindent {\Large December 2003}

\bigskip

\begin{frontmatter}

\journal{Nuclear Instruments and Methods}
\date{}

\title{Start-to-End Simulations of SASE FEL at the TESLA Test
Facility, Phase 1}

\author[DESY]{M.~Dohlus},
\author[DESY]{K.~Fl\"ottmann},
\author[Dubna]{O.S.~Kozlov},
\author[DESY]{T.~Limberg},
\author[Fermilab]{Ph.~Piot},
\author[DESY]{E.L.~Saldin},
\author[DESY]{E.A.~Schneidmiller} and
\author[DESY]{M.V.~Yurkov}

\address[DESY]{Deutsches Elektronen-Synchrotron (DESY),
Notkestrasse 85, 22607 Hamburg, Germany}

\address[Dubna]{Joint Institute for Nuclear Research, Dubna,
141980 Moscow Region, Russia}

\address[Fermilab]{Fermilab, Batavia, IL 60510, USA}

\begin{abstract}
Phase 1 of the vacuum ultra-violet (VUV) free-electron laser (FEL) at the TESLA Test Facility (TTF) recently concluded operation. It successfully demonstrated the saturation of a SASE FEL in the in the wavelength range of 80-120 nm. We present {\it a posteriori} \ start-to-end numerical simulations of this FEL. These simulations are based on the programs Astra and {\tt elegant} for the generation  and transport of the electron distribution. An independent simulation of the intricate beam dynamics in the magnetic bunch compressor is performed with the program CSRtrack. The SASE FEL process is simulated with the code FAST. From our detailed simulations and the resulting phase space distribution at the undulator entrance, we found that the FEL was driven only by a small fraction (slice) of the electron bunch. This ``lasing slice" is located in the head of the bunch, and has a peak current of approximately 3~kA. A strong energy chirp (due to the space charge field after compression) within this slice had a significant influence on the FEL operation. Our study shows that the radiation pulse duration is about 40~fs (FWHM) with a corresponding peak power of 1.5~GW. The simulated FEL properties are compared with various experimental data and found to be in excellent agreement.

\end{abstract}

\end{frontmatter}

\baselineskip 20pt

\clearpage

\section{Introduction}

The vacuum ultra-violet (VUV) free-electron laser (FEL) at the TESLA Test Facility
(TTF), Phase 1 demonstrated saturation in the wavelength range 80-120 nm based on
the self-amplified spontaneous emission (SASE) principle\cite{ttf-sat-desy,ttf-sat-prl,ttf-sat-epj}.
Analysis of experimental data for the radiation properties have led us to the unique conclusion
that SASE FEL produced ultra-short radiation pulse (FWHM duration 30-100~fs)
with GW level of peak power. However the measured properties of the radiation were in strong
disagreement with project parameters for the electron bunch~\cite{ttf-fel-phase1}: we expected
a peak beam current of 500~A, an rms bunch duration of 1~ps, and an energy spread of 0.5~MeV.
Such electron beam parameters would result in operation of the FEL in the saturation regime
only for normalized emittance less or about 2~mm-mrad, while emittance measurements gave values
of approximately 4~mm-mrad for a bunch charge of 1~nC \cite{schreiber-epac02}. Also, the radiation
pulse was by one order of magnitude shorter than the value expected for the aforementioned
project parameters.

Facing this evident disagreement, we did an attempt to reconstruct parameters of the lasing part of
the electron bunch using measured properties of the radiation\cite{ttf-sat-desy,ttf-sat-prl,ttf-sat-epj}.
Actually, radiation measurements were very reliable and accurate, and in combination with
the FEL theory we can infer a lot about the properties of the part of electron bunch that produces
the radiation. In the FEL simulations\cite{ttf-sat-desy,ttf-sat-prl,ttf-sat-epj} lasing part of the
bunch  was approximated by a Gaussian.
A set of parameters for this lasing part of the electron bunch leading to
simulation results, consistent with the radiation measurements, were: a
FWHM bunch duration of 120 fs, a peak current of 1.3~kA, an rms energy
spread of 100~keV, and a rms normalized transverse emittance of 6~mm-mrad. These values,
as used in the FEL-simulation, gave reasonable agreement for the average energy in the
radiation pulse and statistical distributions of the fluctuations of the radiation energy.
However, there were two visible disagreements with experimentally measured radiation
characteristics. First, there was a noticeable difference in the shape of angular distribution
of the radiation intensity. Second, the measured averaged spectrum width was visibly wider
with respect to the simulated one and the spikes in the single-shot spectra were larger in
the experiment than the one simulated.

It is worth mentioning that soon after obtaining the saturation,
attempts for more detailed electron beam measurements were undertaken.
The first one was measurement of the slice energy spread before
compression \cite{schlarb-pac03} which gave the value of about 5~keV.
This measurement clearly indicated that the value of peak current
after compression should be well above 1.3~kA. The second experiment
aimed at direct time-domain measurement with a streak camera \cite{schreiber-epac02} of
the bunch shape and peak current. The results of these measurements
gave $650 \pm100$~fs for rms pulse duration, and 0.6~kA peak current at
3~nC. These results were consistent with another techniques based on a tomographic
reconstruction of the longitudinal phase space\cite{huening-dbi2001}. Both of these
measurement were in fact resolution limited: for instance the temporal resolution of
the streak camera in the experiment reported in \cite{schreiber-epac02} was
200~fs. It was therefore impossible to resolve any fine structure on the bunch charge
density that have characteristic width below $\sim$200~fs. Hence it was impossible
to resolve the lasing fraction of the bunch since it was below 200~fs as we mentioned above.
We should also note that there was no possibility to measure slice emittance and energy
spread at TTF.

Thus there was no complete quantitative description of the TTF SASE FEL operation.
To get a better understanding of the TTF SASE FEL operation we undertook the full physics
simulation of the TTF FEL, Phase 1. This study aimed to trace the evolution of the electron
beam from the photo-cathode to the undulator entrance. Then use the thereby produced
electron distribution to calculate the radiation produced by this electron bunch while
passing in the undulator. In our simulations we tried (to our best knowledge) to reproduce
the main parameters of the machine from the FEL run in September 2001. In some cases, due
to the lack of reliable informations, we had to make simplified assumptions.

Presently there is no  universal particle tracking code capable of the evolution of the
electron beam through the accelerator and bunch compression chain. For the simulation
reported hereafter, the program ASTRA \cite{astra}, which takes into account space charge, but does
not calculate the beam dynamics in the bends, was used in straight transport
sections (gun, capture cavity, accelerating modules, drift spaces). The beam dynamics
in the bunch compressor was independently simulated with the codes {\tt elegant} \cite{elegant}
and CSRtrack \cite{martin} taking into account the effects related to coherent synchrotron
radiation (CSR). Finally, the FEL process was simulated by three-dimensional, time-dependent
code FAST \cite{fast}.

\section{Facility description}

\begin{figure}[b]
\begin{center}
\includegraphics[width=0.9\textwidth]{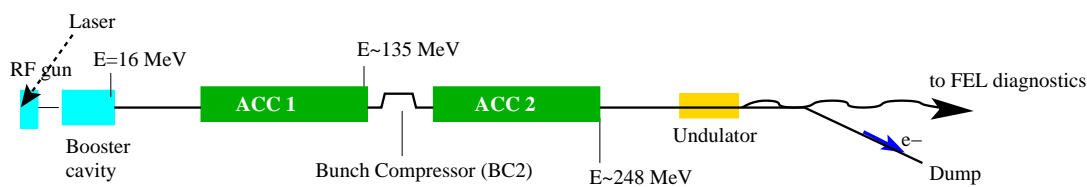}
\end{center}
\caption{\normalsize
Schematic layout of the TESLA Test Facility, Phase 1 }
\label{fig:phil}
\end{figure}

The description of the TTF accelerator (see Fig.~\ref{fig:phil}), operating under standard lasing
conditions, can be found in\cite{ttf-sat-prl,ttf-sat-epj} and references therein.
The RF gun consists of an L-band cavity (1+1/2 cell) incorporating a CsTe$_2$ photo-cathode
illuminated by a UV laser with a (Gaussian) pulse distribution of 7-8~ps rms. The electron bunch
with the charge 2.8 nC and energy about 4~MeV is extracted from the gun (nominal laser launch phase
is 40 deg) and is then accelerated in the booster cavity up to 16~MeV.  The phase of the
booster cavity is normally chosen such that the total (correlated) energy spread is minimized.
Passing a rather long drift, the beam is then injected into the superconducting TESLA module (ACC1)
where it is accelerated up to 135 MeV off-crest to impart the proper correlated energy spread for subsequent
compression in the following four-bend magnetic chicane (BC2). Without compression the bunch is rather
long (about 3.5 mm rms - longer than the laser pulse due to Coulomb repulsing in the injector). Thus,
at the nominal compression phase (10
deg off-crest) the bunch accumulates RF curvature leading to a "banana"
shape on the longitudinal phase plane after compression. The resulting
bunch shape in time domain constitutes a short high-current leading
peak and a low-current long tail. After compression the beam passes a 5~m long drift, the
second TESLA module ACC2 (being accelerated up to 248
MeV), and another drift (about 20 m) that includes a collimation
section and a transverse matching section. Finally the beam enters the
undulator consisting of three 4.5 m long modules where a short SASE
pulse (wavelength below 100 nm) is produced by the leading peak of the
bunch. The electron beam is separated from the photon beam thanks to the spectrometer dipole,
and bent in a beam dump while the photon beam goes to the photon diagnostic area.

\begin{figure}[b]
\begin{center}
\includegraphics[width=0.8\textwidth]{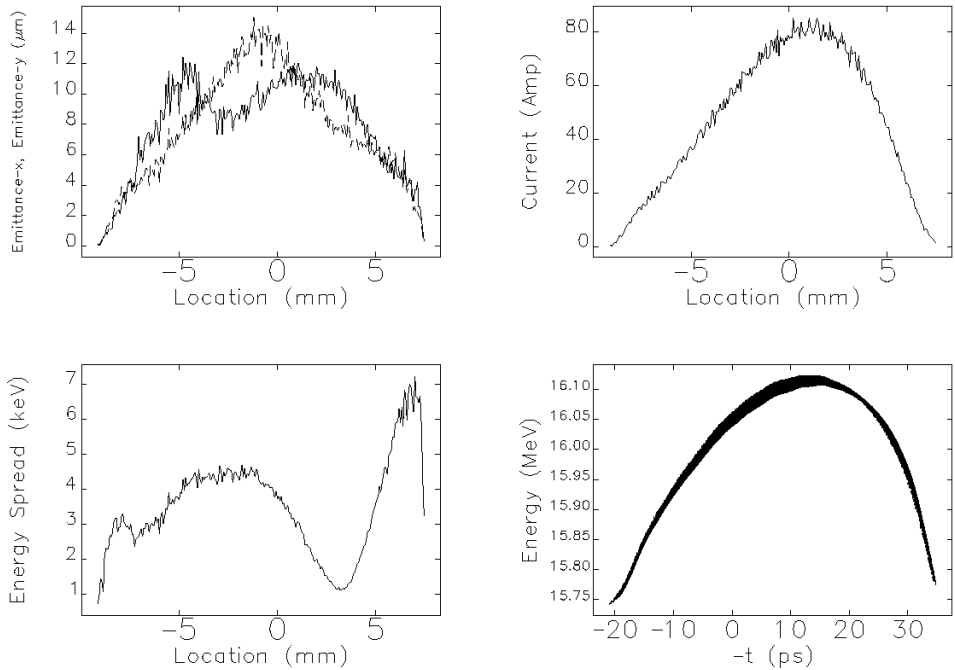}
\end{center}
\caption{\normalsize
Normalized slice emittance (x - line, y - dash), current, slice energy
spread, and longitudinal phase plane at the entrance to ACC1. Bunch
head is on the right
}
\label{fig:acc1}
\end{figure}

\section{Simulations of the beam dynamics in the accelerator}

The initial part of the accelerator, from the photo-cathode to BC2 entrance,
was simulated with Astra \cite{astra}, a program that includes space
charge field using a cylindrical symmetric grid algorithm. The beam was
then tracked through the bunch compressor with {\tt elegant}~\cite{elegant} that includes
a simplified model (based on a line charge approximation) of the CSR wake\cite{we-csr,stup-csr}.
Downstream of BC2, Astra was again used up to the undulator entrance because the space charge
induced-effects are significant for the strongly compressed part of the bunch. SASE FEL
process in the undulator was simulated with three-dimensional time-dependent code FAST \cite{fast}.
In order to check that we did not miss any important CSR-related effects in the bunch compressor, we
performed an independent simulations of the bunch compressor with a newly developed
code CSRtrack \cite{martin}. This latter code incorporates a two-dimensional self-consistent model
of the beam dynamics.

\begin{figure}[p]

\begin{center}
\includegraphics[width=0.8\textwidth]{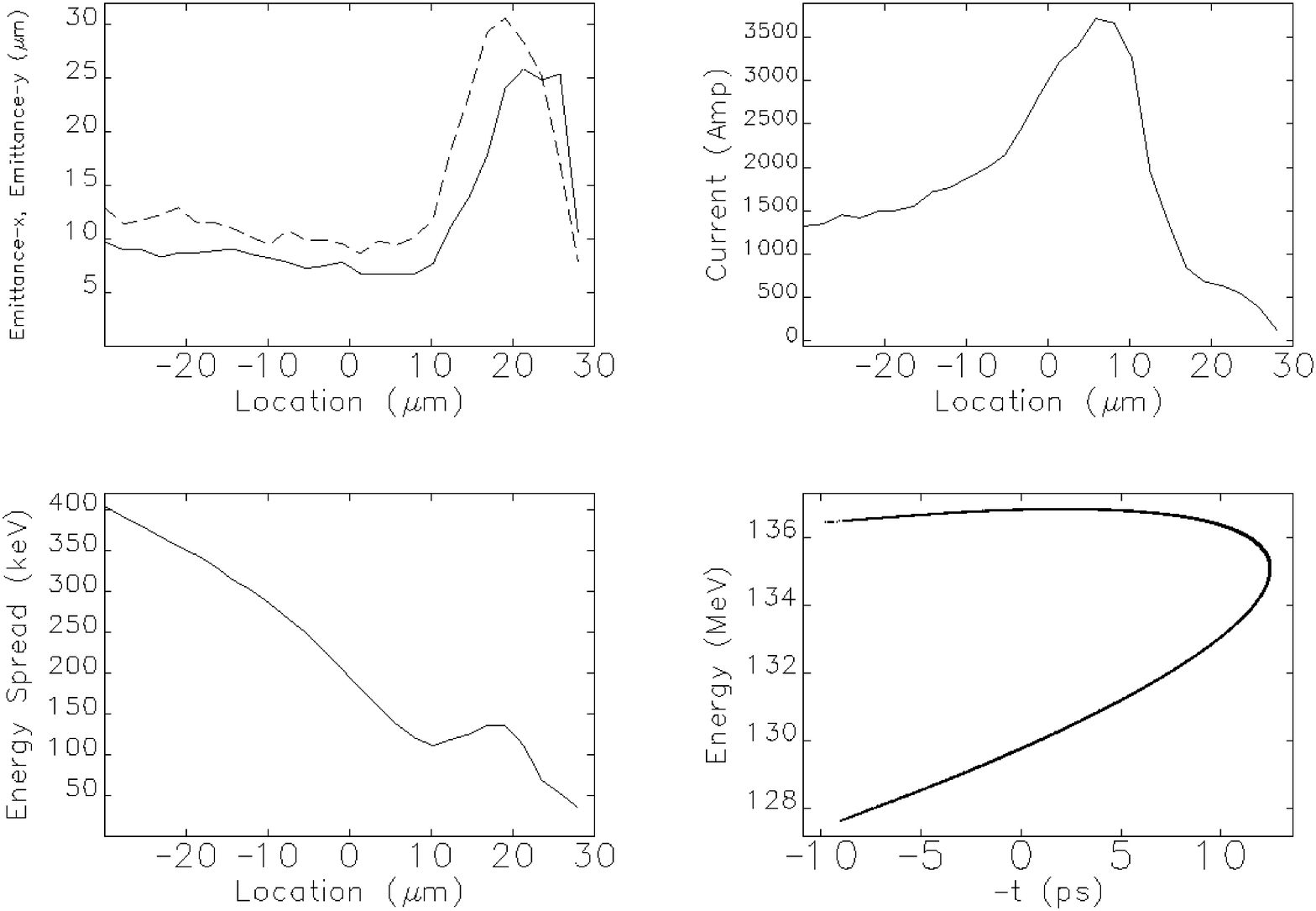}
\end{center}
\caption{\normalsize
Normalized slice emittance (x - line, y - dash), current, slice energy
spread in the front part of the bunch, and longitudinal phase plane.
The position is behind the bunch compressor.
CSR is off.
Bunch head is
on the right
}
\label{fig:bc2-csr}

\begin{center}
\includegraphics[width=0.8\textwidth]{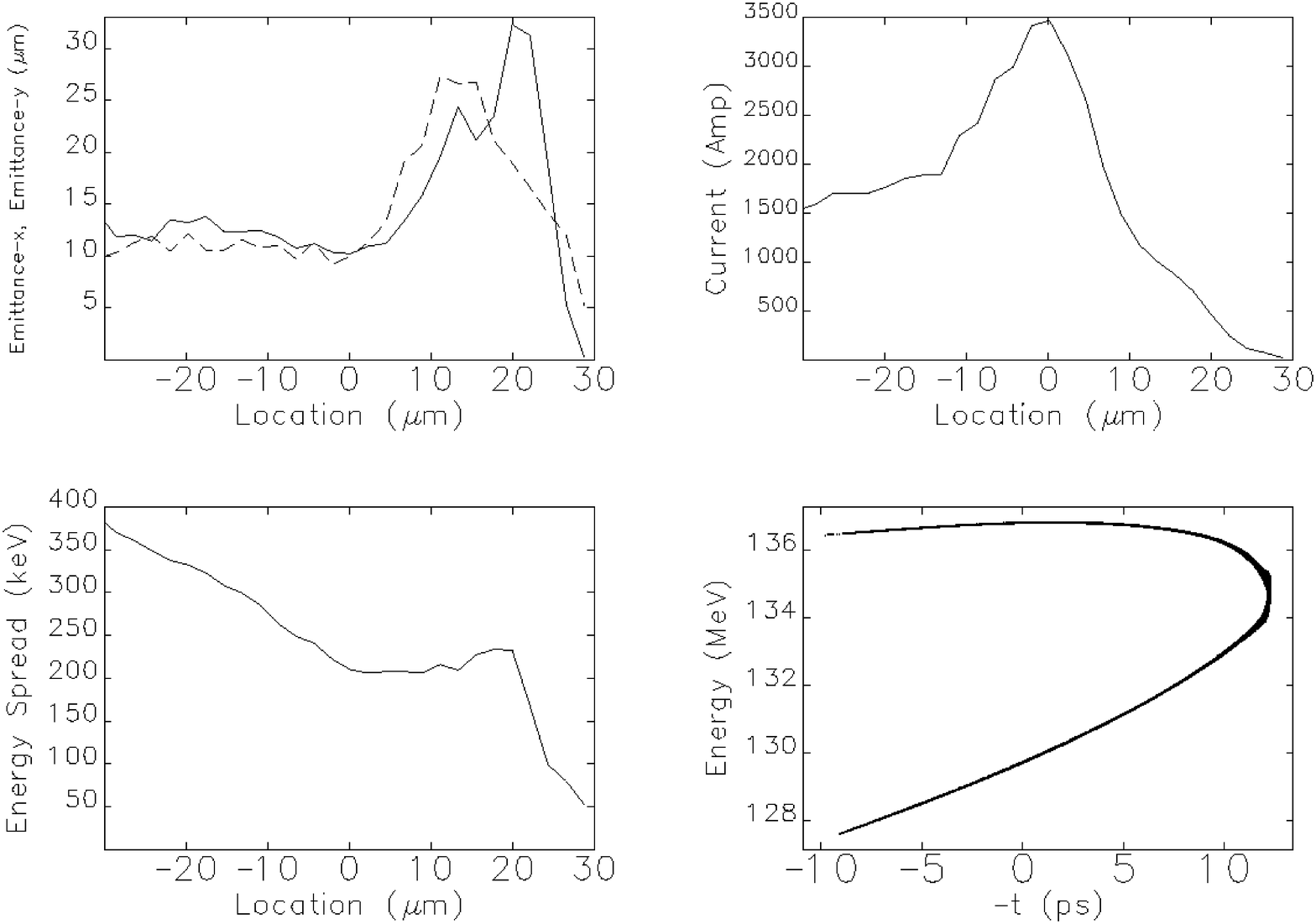}
\end{center}
\caption{\normalsize
Normalized slice emittance (x - line, y - dash), current, slice energy
spread in the front part of the bunch, and longitudinal phase plane.
CSR is on.
The position is behind the bunch compressor. Bunch head is
on the right
}
\label{fig:bc2}

\end{figure}

\begin{figure}[tb]

\begin{center}
\includegraphics[width=0.8\textwidth]{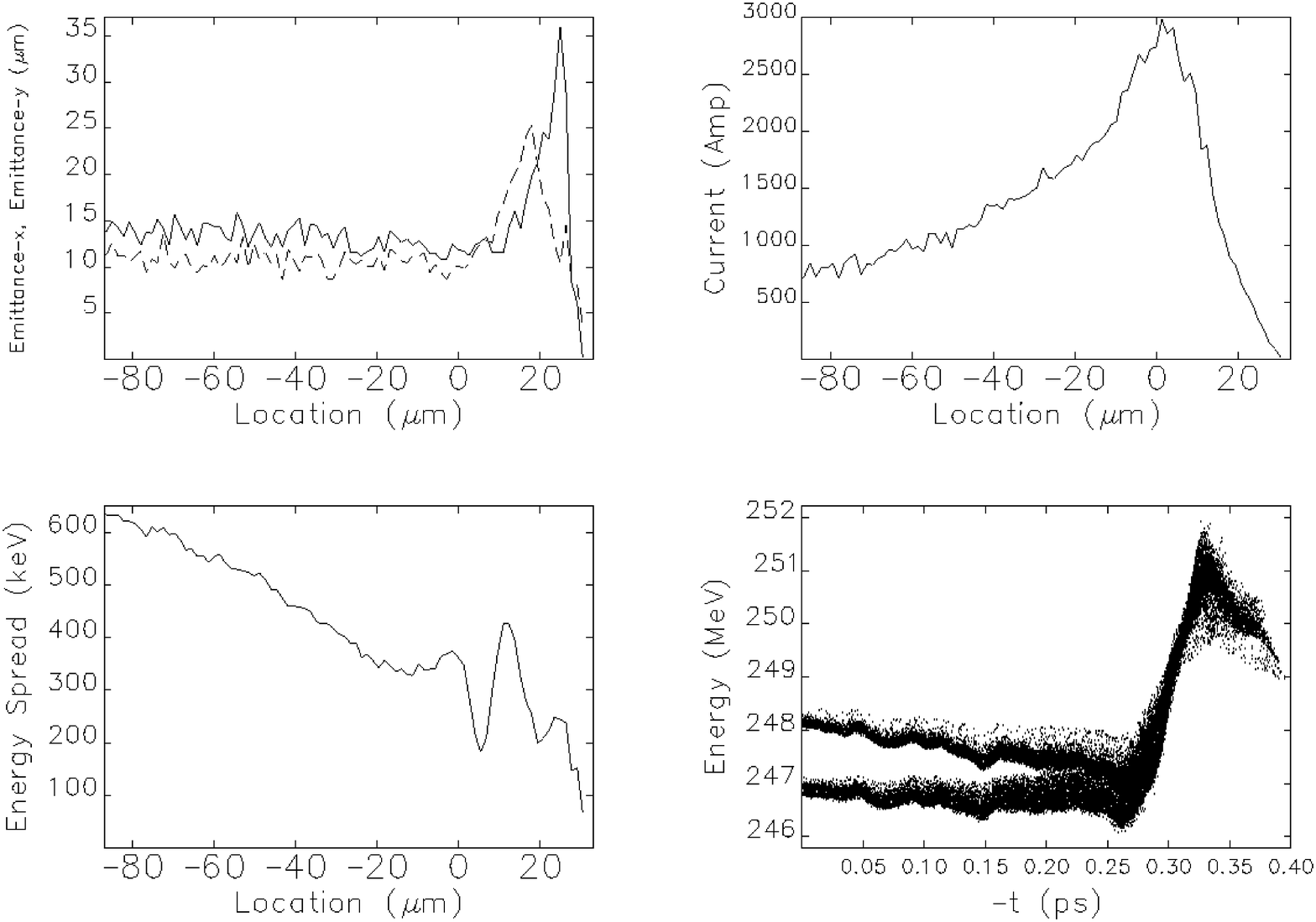}
\end{center}
\caption{\normalsize
Normalized slice emittance (x - line, y - dash), current, slice energy
spread, and longitudinal phase plane
for the front part of the bunch. The position is at the undulator
entrance. Bunch head is on the right
}
\label{fig:und-ent}
\hspace*{1cm}
\end{figure}

The Astra simulation from the cathode to the bunch compressor was performed
using $2\times 10^5$ macro-particles. The main parameters of the electron beam,
calculated for various longitudinal slices along the bunch, at the entrance to
ACC1 are presented in Fig.~\ref{fig:acc1}. The chosen bin size is such that the
energy chirp does not contribute to the slice energy spread.  The latter parameter
is very important since it defines the width of the leading peak and the peak current
after compression. One can see that it varies along the
bunch, taking the value of about 3.5 keV at the location $s=0$ (this part of
the bunch is then put into a local full compression in BC2). The resulting values for the
local energy spread in Fig.~\ref{fig:acc1} are in agreement with the
measurements \cite{schlarb-pac03}.  The main source of the build-up of the local energy
spread (as we see it from this simulation) is a transverse variation of the longitudinal
space charge field in the injector. Since it is a coherent effect (a particle's energy deviation
is correlated with its position in the bunch, in particular, with its transverse offset within
a given slice), the frequently used notions of "uncorrelated" or "incoherent" energy spread
are not adequate here. As we will see, the correlations can be important when we compress the
beam.

At the entrance of BC2 the output distribution of macro-particles is converted into the input
distribution for {\tt elegant}. It is then tracked through the bunch compressor
($R_{56} \simeq 23$ cm) without CSR wake included.
The resulting distribution on longitudinal phase plane and slice
parameters for the leading peak are presented in Fig.~\ref{fig:bc2-csr}. One can
see that the peak current and the width of
the current spike are in a reasonable agreement
with analytical estimate, presented in Appendix.
The difference can be explained by the non-Gaussian energy distribution in a slice
before compression and by the fact that the beam entered ACC1 with the
energy 16 MeV and some energy chirp, etc. Also, important correlations
in particle distribution are not considered in Appendix. What
requires some explanations is a behavior of the slice emittance in
Fig.~\ref{fig:bc2}. As we have already mentioned, the main effect, responsible for a
slice energy spread in the injector, is the transverse variation of the
longitudinal space charge field. An energy offset of a particle in a
slice (with respect to on-axis particle) is correlated with its
transverse offset (and, since particles are moving transversely, with
a betatron amplitude). A sign of the energy offset depends on the
position of the slice in the bunch.  Indeed, the head of the bunch is
accelerated due to space charge field, so that for a given slice a
particle with a finite betatron amplitude gets less energy than an
on-axis particle. The situation is opposite for the slices which are
on the left from a minimum energy spread in Fig.~\ref{fig:acc1}: the larger
transverse offset, the larger positive energy deviation. During
compression particles with higher energies (in our case, with larger
betatron amplitudes) move forward, this explains the large slice emittance values
for the slices in the leading front of the spike. On the other hand,
due to this effect a slice with the maximal current has less "bad"
particles, so that the emittance there is visibly less than that at $\mathrm{s} \simeq 0$
in Fig.~\ref{fig:acc1}. We also tried to put the head of the bunch (say
$\mathrm{s} \simeq 5$ \ mm in Fig.~\ref{fig:acc1}) into a local full
compression.  The result was that after compression emittance was
smoothly decreasing in the leading front of the spike. It is worth noting that when
we prepared a 6-D Gaussian bunch with no correlations (at the entrance to ACC1) and
track it through BC2, the slice emittance after compression was the same in all slices and
was equal to its value before compression.

Our next step was to track the same particle distribution through the
bunch compressor, taking into account CSR wake in a simplified model
used in {\tt elegant}. The beam parameters behind BC2 are
presented in Fig.~\ref{fig:bc2-csr}. Note that due to CSR-induced effects the peak current slightly
decreased, by less than 10\%.  Slice emittance (in horizontal plane)
in the slice with maximal current increased by some 50\%, and the
local energy spread by almost a factor of 2.

It looks surprising that the peak current is almost unchanged
in the presence of the CSR wake. Indeed without including CSR, the final shape of
the bunch forms in the end of the third - begin of the fourth dipole of BC2. If one does a
naive estimate for the energy kicks due to CSR (on the way to the end of the fourth dipole)
for such a narrow peak with so high current, and applies $R_{56}$ to the end of compressor,
then one finds out that the distribution of current should be strongly disturbed. The reason why
this does not happen can be explained as follows.  For our range of parameters, one can not
neglect coupling between transverse and longitudinal phase spaces in the bunch compressor (the
importance of this effect was pointed out in studies of CSR microbunching
instability \cite{we-micr,stup-micr,kim-micr}), described by linear transfer
matrix elements $R_{51}$ and $R_{52}$ (the net effect through the whole
compressor is zero to first order, and higher order terms are negligible). This coupling makes
the leading peak effectively much longer and the current much lower than in the case of zero
emittance (the spike gets cleaned up only at the very end of the fourth dipole).
Thus, the energy kicks due to CSR are strongly suppressed, and the current spike survives.

In order to confirm this result and to check that we did not miss any important effect, using a
simplified CSR model in {\tt elegant}, we performed alternative simulations for the bunch compressor
with the "first principles" code CSRtrack. For technical reasons, the incoming distribution of
macro-particles has to be simplified in those simulations (for instance, the above mentioned
correlations were neglected), so that slice parameters after compression are not exactly reproduced
even without CSR.  Nevertheless, the main results of simulations with {\tt elegant} were
confirmed: the peak current decreases by less than 10\%, and the slice emittance
growth is in the range of several tens of per cent.

>From BC2 exit to the undulator entrance the tracking was done with Astra. The reason is that a
simple estimate predicts very strong longitudinal space charge effect in the leading peak. Indeed,
for a Gaussian bunch with an rms length $\sigma_z$ and a peak current $I$ the change of the
peak-to-peak energy chirp $\Delta \gamma$ (in units of the rest energy) in a drift can be estimated
as

\begin{displaymath}
\frac{d (\Delta \gamma)}{d z} \simeq \ 2.4 \ \frac{I}{I_{\mathrm{A}}} \
\frac{\ln (\gamma \sigma_z/\sigma_{\bot})}
{\sigma_z \gamma^2}
\end{displaymath}

\noindent where $I_{\mathrm{A}} = 17 $kA is the Alfven current, $\gamma$ the
relativistic factor and $\sigma_{\bot}$ the rms transverse size of the beam.
This formula holds when $\sigma_z \gamma \gg \sigma_{\bot}$. The estimate shows that for
the leading peak the energy chirp should be in the range of several MeV.

To reduce numerical calculation effort, we did not track the entire distribution since
we were anyway not interested in the parameters of the long low-current tail. So, we cut
the tail away and tracked particles in the head of the bunch (typically this part was a
few hundred $\mu$m long in our simulations). The parameters of the front part of the bunch
at the undulator entrance are shown in Fig.~\ref{fig:und-ent}.
One can notice a big energy chirp
due to the space charge within the current spike. Note also that due to
Coulomb repulsing the spike gets wider, and the peak current decreases
by approximately 20\%.  A change of the local energy spread is due to
transverse variations of the longitudinal space charge field. As one
would expect, the local minimum of the energy spread is close to the position of
maximal current (where derivative of the current is zero).

\section{Simulation of the SASE FEL process}

\begin{figure}[t]

\includegraphics[width=0.5\textwidth]{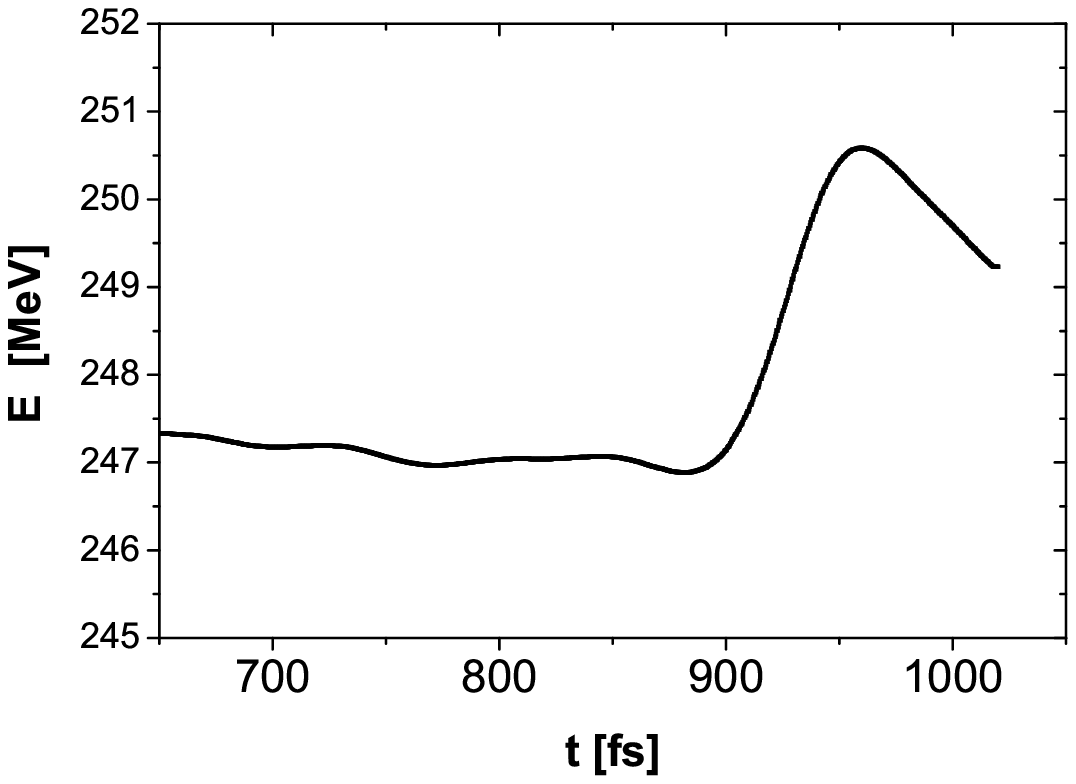}
\includegraphics[width=0.5\textwidth]{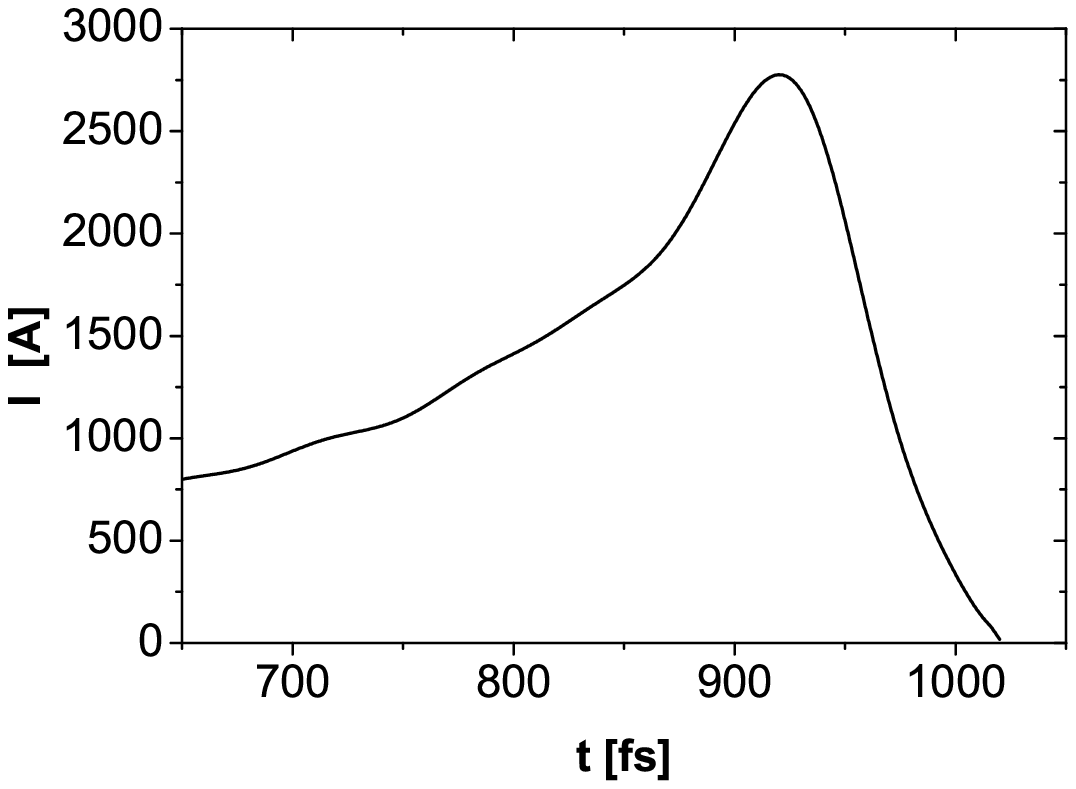}

\vspace*{-5mm}

\includegraphics[width=0.5\textwidth]{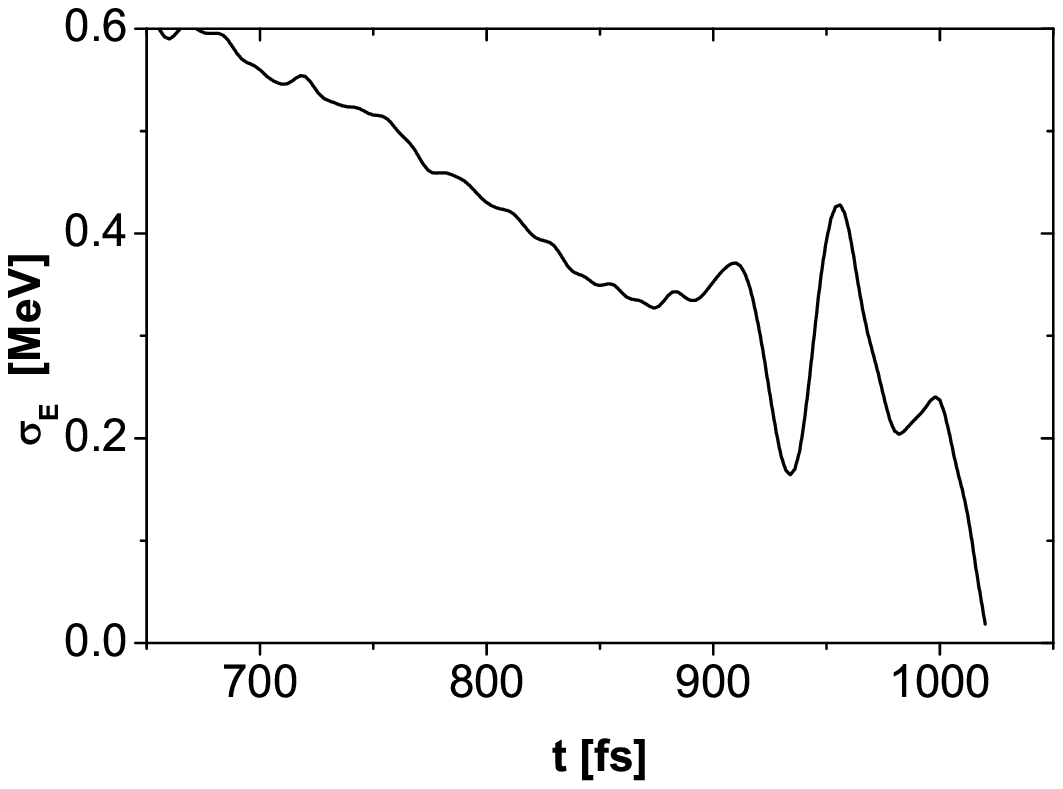}
\includegraphics[width=0.5\textwidth]{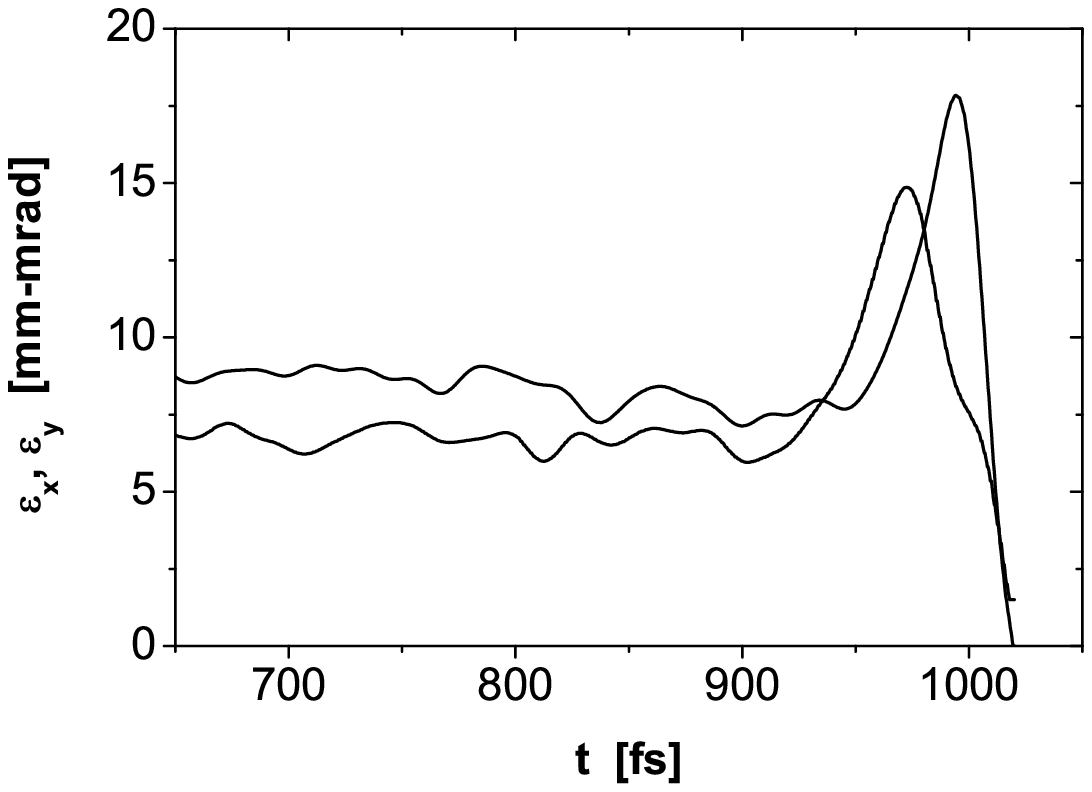}

\vspace*{-3mm}

\caption{\normalsize Mean energy,
current, slice emittance and slice
energy spread along the bunch at the undulator entrance.
Bunch head is at the right side
}
\label{fig:bunch-ue}
\end{figure}

The simulations SASE FEL process were performed with the three-dimensional
time-dependent FEL code FAST \cite{fast}. Before describing the results of
these simulations we should discuss a difficulty, connected with the
data transfer from a beam dynamics simulation code to an FEL simulation
code. While the data exchange
between beam dynamics simulation codes (for instance Astra $\to$
{\tt elegant} and {\tt elegant} $\to$ Astra) is straightforward and technically
simple, a direct loading, for instance, of Astra output distribution
as FAST input distribution is impossible. The reasons for this are:
completely different time scales of the processes, and a
necessity to avoid an artificial noise of macro-particles in FEL
spectral range (in addition, in case of a SASE FEL simulation it is
also necessary to correctly simulate a real shot noise in the beam - an
input signal for SASE FEL).

The standard way of preparing input data for an FEL code is as follows:
a macro-particle distribution at the undulator entrance is cut into
longitudinal slices. A mean energy, rms energy spread, current,
rms emittances etc. are calculated for each slice.
Then regular distributions (normally - Gaussian) in each slice are
used in FEL code for energy and both transverse phase spaces, having
the same mean and rms values as input distribution.
Clearly, since only mean and rms values are extracted from the original distributions,
an essential information can be lost if the distributions are strongly non-Gaussian.
For energy distribution it is sufficient to know only mean and rms values if the rms
value is visibly less than FEL parameter $\rho$ \cite{bonifacio,book},
which is often the case. For transverse phase space, however, such a simplified
procedure may not be satisfactory, especially when the rms emittance is
strongly influenced by "halo" particles. More adequate procedure could
be a two-dimensional Gaussian fit (with the least-square method) for
each longitudinal slice of each phase plane. Indeed, not only first and second, but
also higher-order momenta of the distribution contribute to the fit
result. A disadvantage of this procedure is the requirement to have a
lot of macro-particles in each slice in order to do a reliable fit.
To avoid this difficulty we did the following. We used Gaussian fit in
$x$ and $y$-phase spaces for the part of the bunch shown in Fig.~\ref{fig:und-ent}, without
cutting it into slices, and got new values for rms emittances. Then the
slice emittance curves were rescaled in accordance with the fit result. Finally, after some smoothing
we got the input parameters for FAST which are shown in Fig.~\ref{fig:bunch-ue}.
In the simulation we assumed that all slices are perfectly
matched to the undulator, and all the centroids are on the ideal orbit.

We performed 300 statistically independent runs with FEL simulation
FAST. Simulation results are compared with experimental data published
earlier \cite{ttf-sat-desy,ttf-sat-prl,ttf-sat-epj,fel2002-stat}.
Despite experimental data were collected during extended period of
time, for the above mentioned publications we selected only those data
which corresponded to similar tuning of the accelerator. The same
settings of the accelerator were used in the start-to-end simulations.

\subsection{Radiation energy and fluctuations}

\begin{figure}[b]

\includegraphics[width=0.5\textwidth]{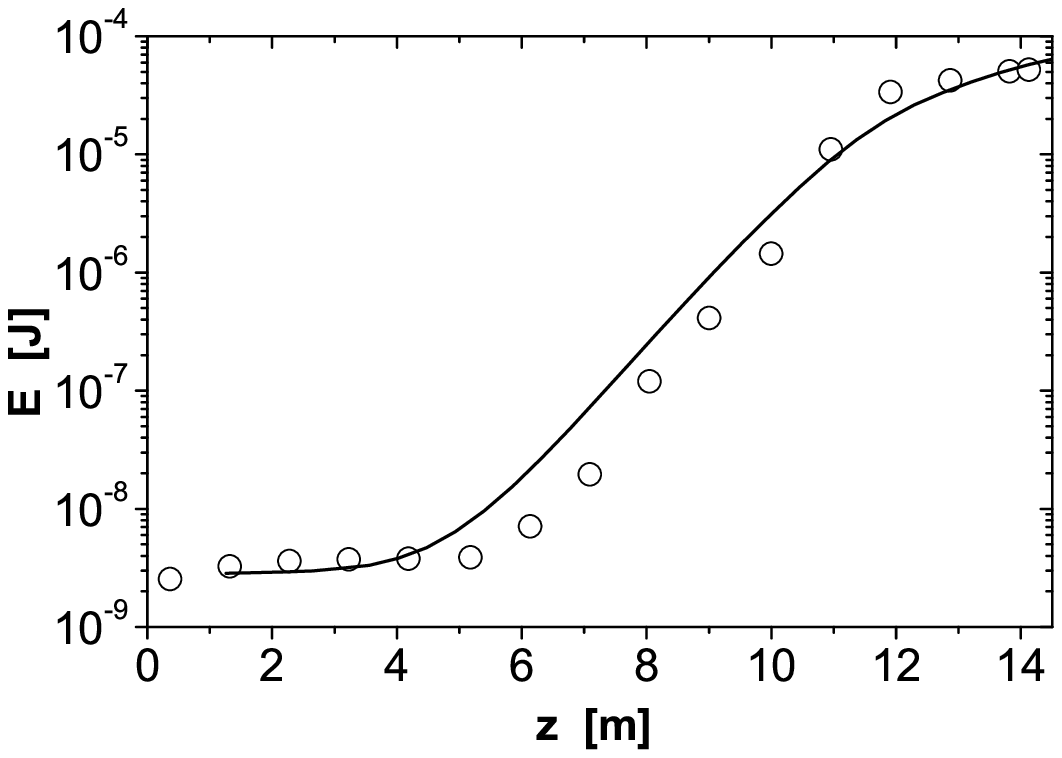}

\vspace*{-59mm}

\hspace*{0.5\textwidth}
\includegraphics[width=0.5\textwidth]{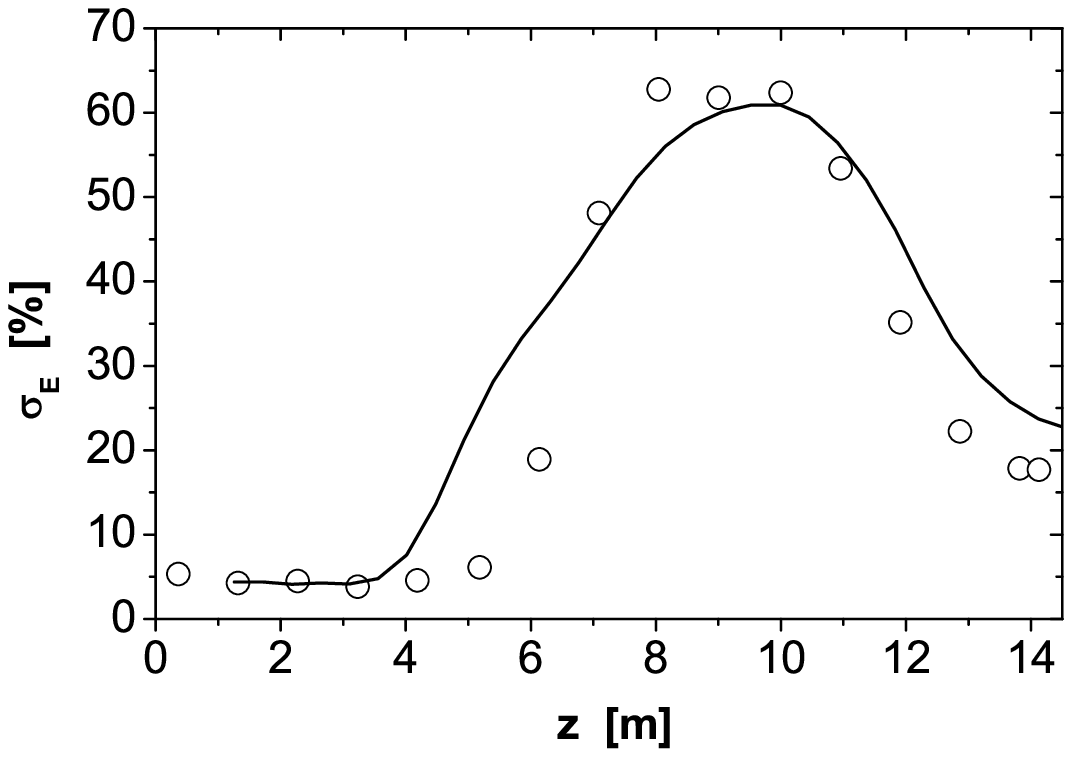}

\caption{\normalsize Energy in the radiation pulse (left plot) and
fluctuations of the energy in the radiation pulse (right plot) versus
undulator length.
Circles represent experimental data \cite{ttf-sat-prl}. Solid lines
represent simulation results with code FAST using bunch
parameters shown in Fig.~\ref{fig:bunch-ue}
}
\label{fig:ez}

\includegraphics[width=0.5\textwidth]{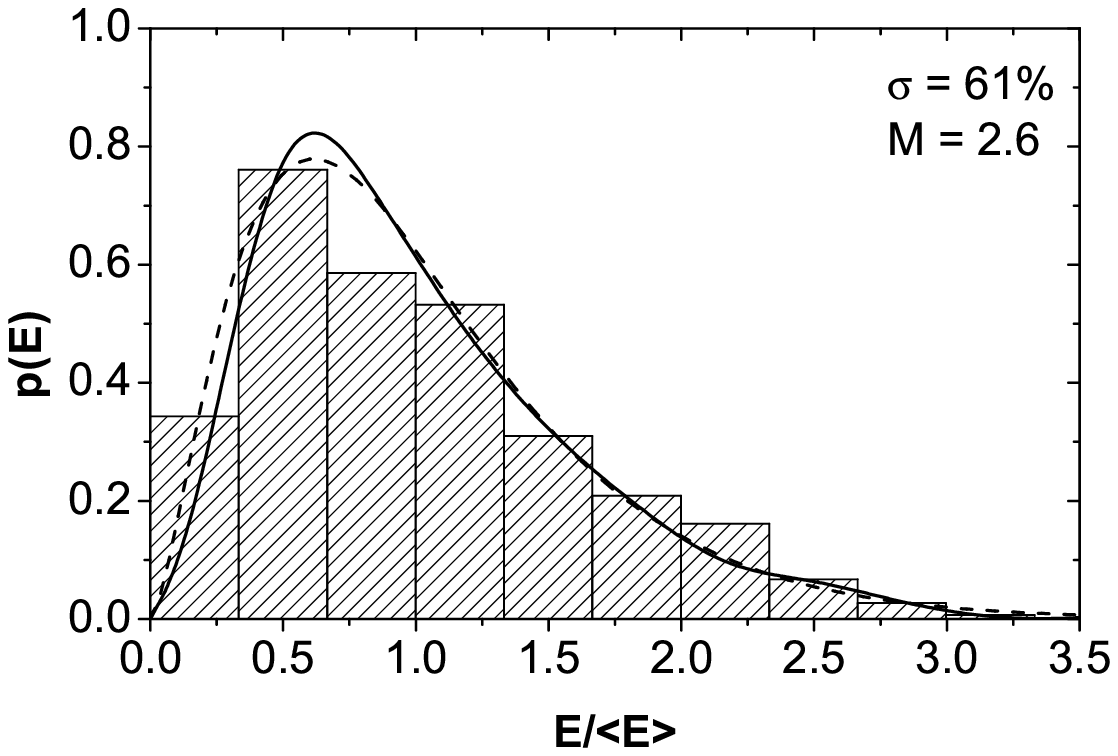}

\vspace*{-59mm}

\hspace*{0.5\textwidth}
\includegraphics[width=0.5\textwidth]{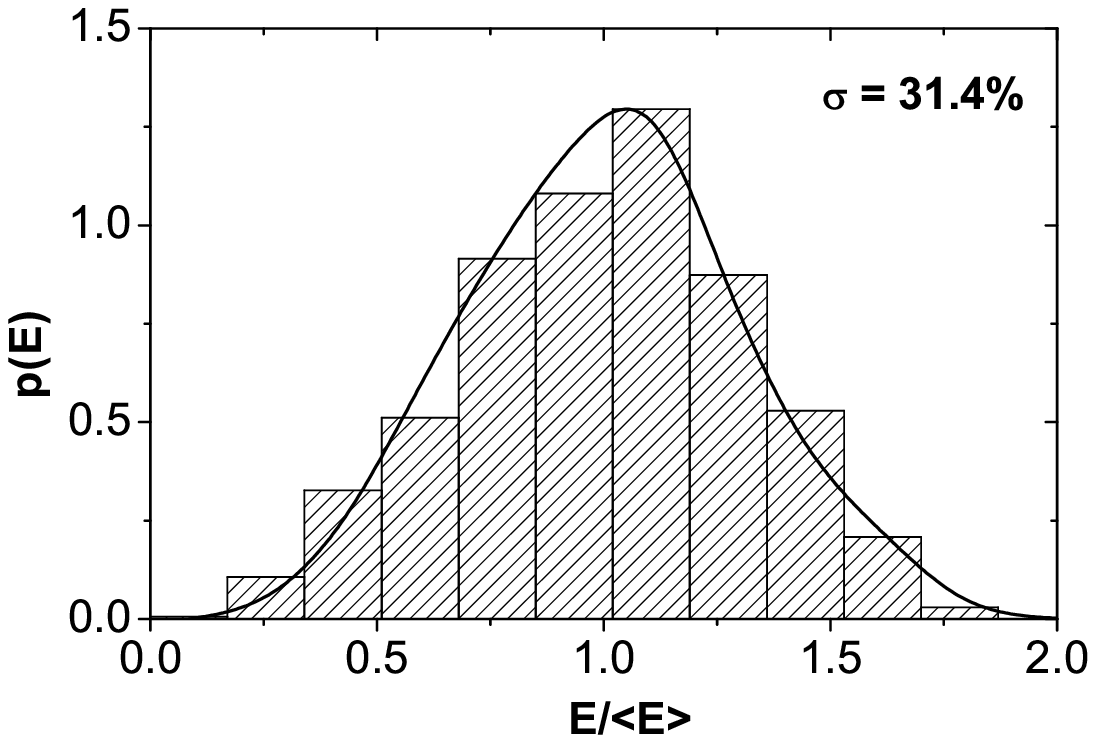}

\caption{\normalsize
Probability distributions of the energy in the radiation pulse for
linear (left plot) and nonlinear regime (right plot) regime. Bars
represent experimental data \cite{fel2002-stat}. Solid lines
represent simulations with code FAST. Dashed line in the left plot
shows gamma distribution with the parameter M = 2.6 } \label{fig:he3}

\end{figure}

Left plot in Fig.~\ref{fig:ez} presents the average energy in the
radiation pulse versus undulator length. Details of experiment are
presented in \cite{ttf-sat-prl}. The interaction length has been
changed by means of switching on electromagnetic correctors installed
inside the undulator. The value of the orbit kick provided by a
corrector was sufficient to stop FEL amplification process downstream
the corrector. The radiation energy has been measured by means of an
MCP-based detector of 10~mm diameter installed 12~m downstream the
undulator \cite{mcp-detector}. When the FEL interaction is suppressed
along the whole undulator length, the detector shows the level of
spontaneous emission of about 2.5~nJ collected from the full undulator
length. In order to reproduce correctly experimental situation, the
simulation results were distorted with the same level
of noise (5\%) as that provided by the radiation detector. Comparison
of experimental and simulation results shows reasonable agreement which
is within limits of accuracy of experiment. Experimentally measured
power gain length is about 70~cm. Calculations of power gain length for
the same electron beam without energy chirp give the value which is almost twice
shorter, about 40~cm.
Explanation of this puzzle is in strong suppression of the FEL gain by
the energy chirp in the electron bunch, about 1\%  on a scale of
cooperation length (to be compared with the FEL parameter $\rho$ \cite{bonifacio,book}
which is about 0.5\%).

Each circle in the left plot in Fig.~\ref{fig:ez} is the result of averaging over
100 shots. The energy in the radiation pulse fluctuates from shot to
shot. The plot for the standard deviation $\sigma$, is presented on the
right side in this Figure. At the initial stage fluctuations are
defined mainly by the fluctuations of the charge in the electron bunch
and the accuracy of measurements. When the FEL amplification process takes
place, fluctuations of the radiation energy are mainly given by the
fundamental statistical fluctuations of the SASE FEL radiation
\cite{statistics}.  A sharp drop of the fluctuations in the last part
of the undulator is a clear physical confirmation of the saturation
process. Detailed measurement of probability distributions were made
for the end of the linear regime (undulator length 9~m), and in the
nonlinear regime. A comparison of experimental and simulated probability
distributions is presented in Fig.~\ref{fig:he3}. It is seen, as expected
from theory\cite{statistics}, that in the linear regime both distributions are
gamma-distributions

\begin{displaymath}
p(E) = \frac{M^M}{\Gamma (M)}
\left( \frac{E}{\langle E\rangle }\right)^{M-1} \frac{1}{\langle
E\rangle } \exp \left( -M \frac{E}{\langle E\rangle } \right) \ ,
\label{gamma}
\end{displaymath}

\noindent with the value $M = 1/\sigma ^2 \simeq 2.6$ (corresponding
to $\sigma = 62$\% at the undulator length 9~m, see Fig:\ref{fig:ez}).
Right plot in Fig.~\ref{fig:he3} shows comparison of the
measured and simulated statistical distributions for the nonlinear
regime which again are in excellent agreement.

\subsection{Angular divergence}

\begin{figure}[h]

\includegraphics[width=0.5\textwidth]{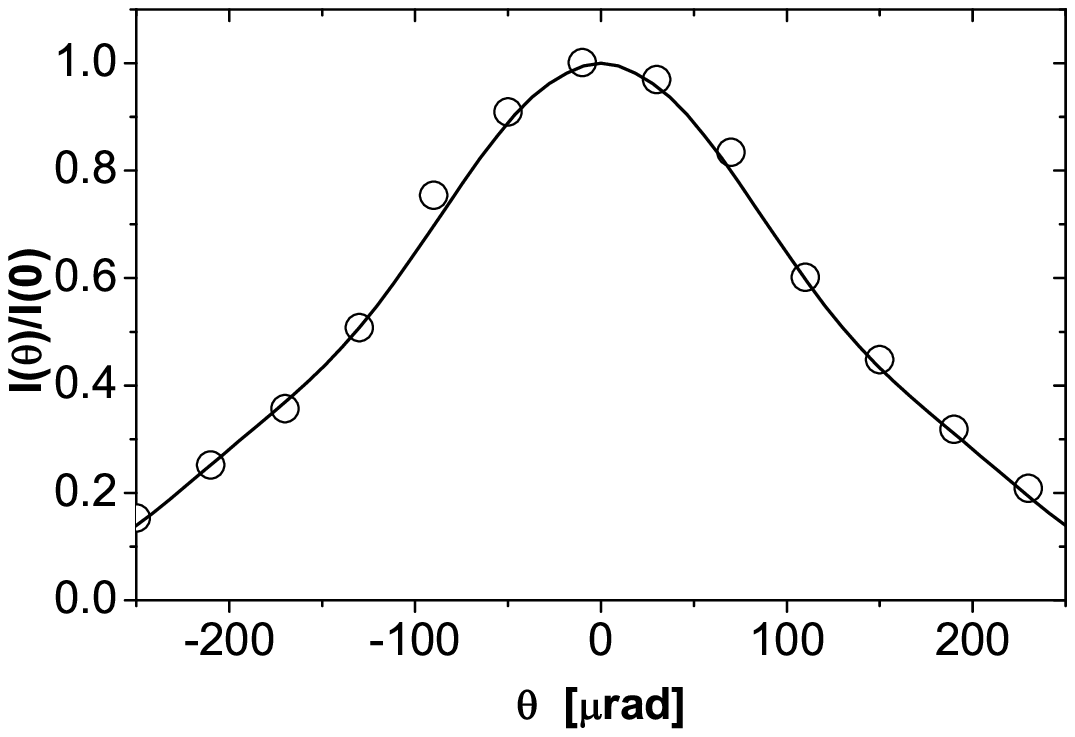}

\vspace*{-57mm}

\hspace*{0.5\textwidth}
\includegraphics[width=0.5\textwidth]{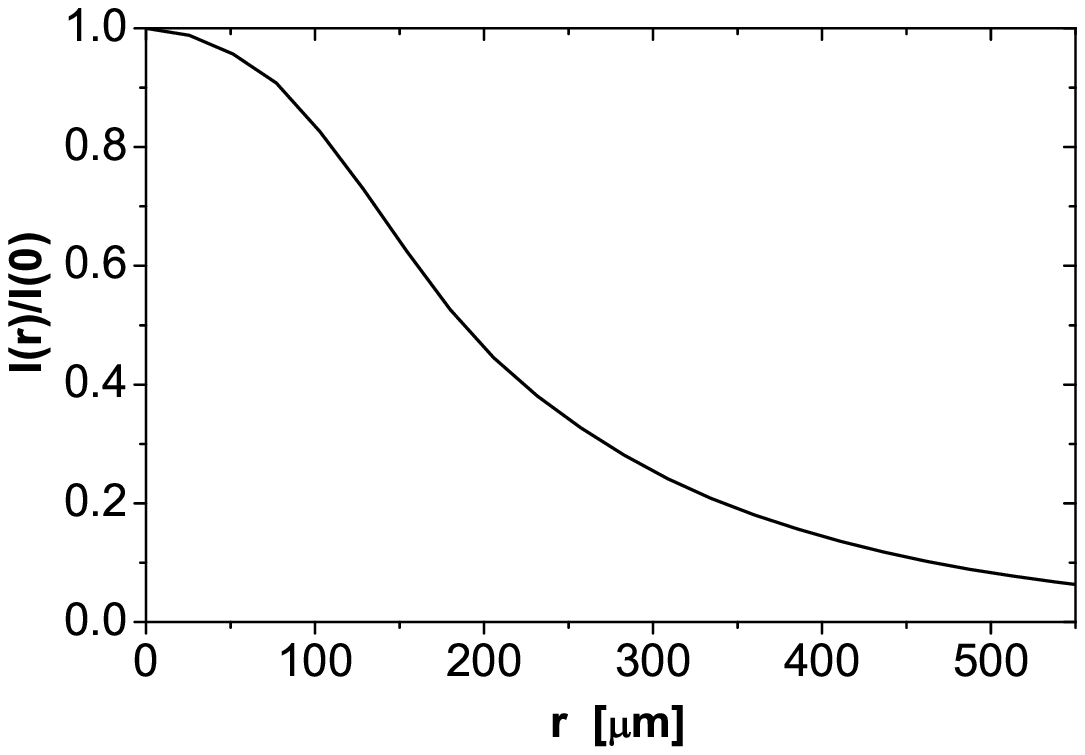}

\caption{\normalsize
Angular distribution of the radiation intensity in the far zone (left
plot) and in the near zone (right plot). TTF FEL operates in the
nonlinear regime. Circles represent experimental data
\cite{ttf-sat-prl}, and solid curves show results of simulations with
code FAST
}
\label{fig:dir}
\end{figure}

Figure~\ref{fig:dir} shows the angular divergence of the FEL radiation.
Measurements were performed by means of scanning a 0.5~mm aperture across
the radiation beam and recording the transmitted intensity on a downstream detector
\cite{ttf-sat-prl}. The simulated curve has been produced by recalculating
the radiation field from the near (undulator exit) into the far zone.
The agreement between simulation and measurement is perfect: even fine details
in the distribution are well reproduced. We can thus state that our simulation
model provides the same field distribution at the undulator exit as it was in the
experiment. We should emphase that the disagreement in angular distribution presented in
a previous analysis \cite{ttf-sat-prl} was a strong indication on rather complicated
physical process. Experimental procedure was very reliable, since it was simple relative
measurement of the intensity with respect to maximum. Indeed, we can conclude that long
spanning tails in the angular distributions is consequence of the strong longitudinal
energy chirp which significantly disturbs the beam radiation mode. Experimentally it was
not possible to measure spot size of the radiation at the undulator exit. Right plot in
Figure~\ref{fig:dir} shows relevant distribution reconstructed from the simulation data.
Figure~\ref{fig:dir} led us to the conclusion that there should be a high degree of
transverse coherence in the FEL radiation, as it was proven in a dedicated experiment at
TTF FEL \cite{rasmus}.

\begin{figure}[b]

\hspace*{0.02\textwidth}
\includegraphics[width=0.46\textwidth]{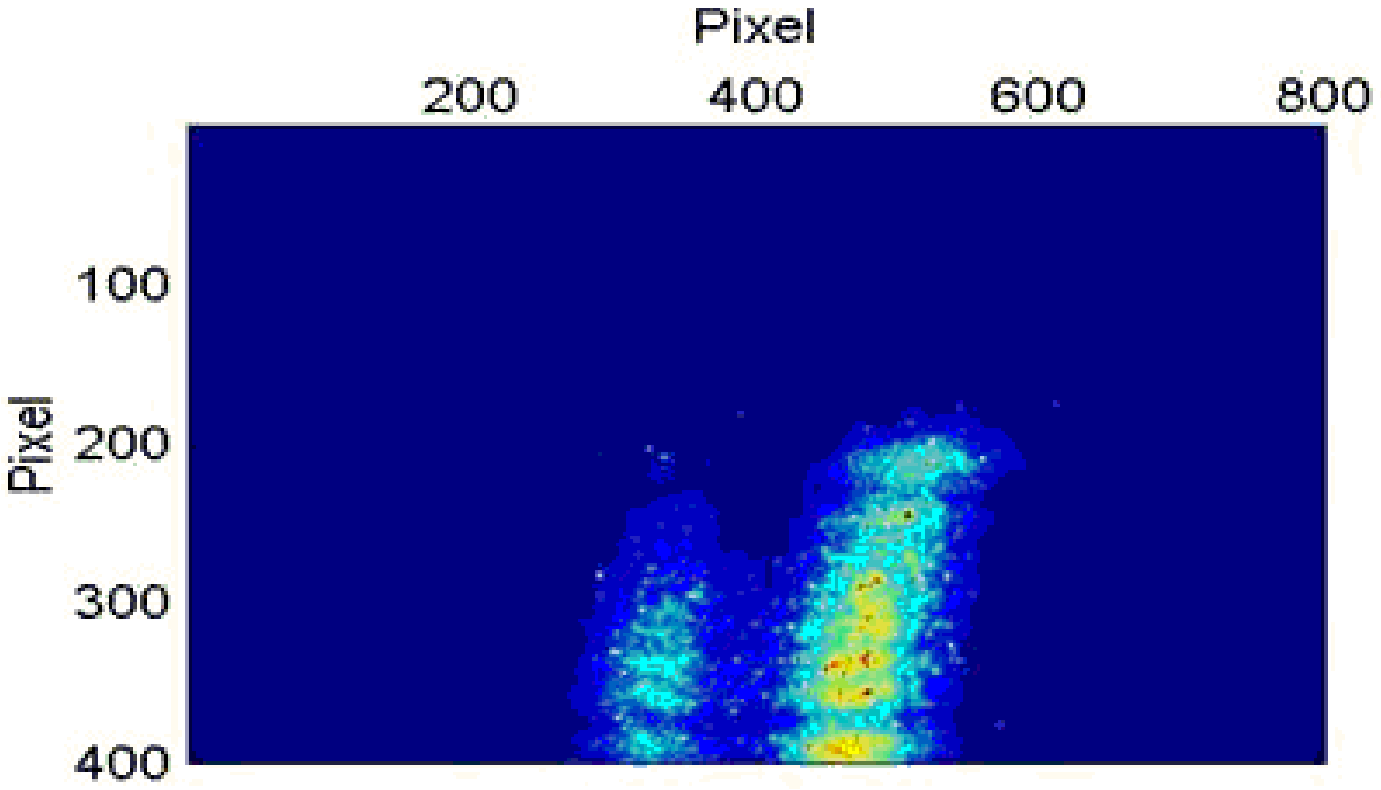}

\vspace*{-44mm}

\hspace*{0.575\textwidth}
\includegraphics[width=0.44\textwidth]{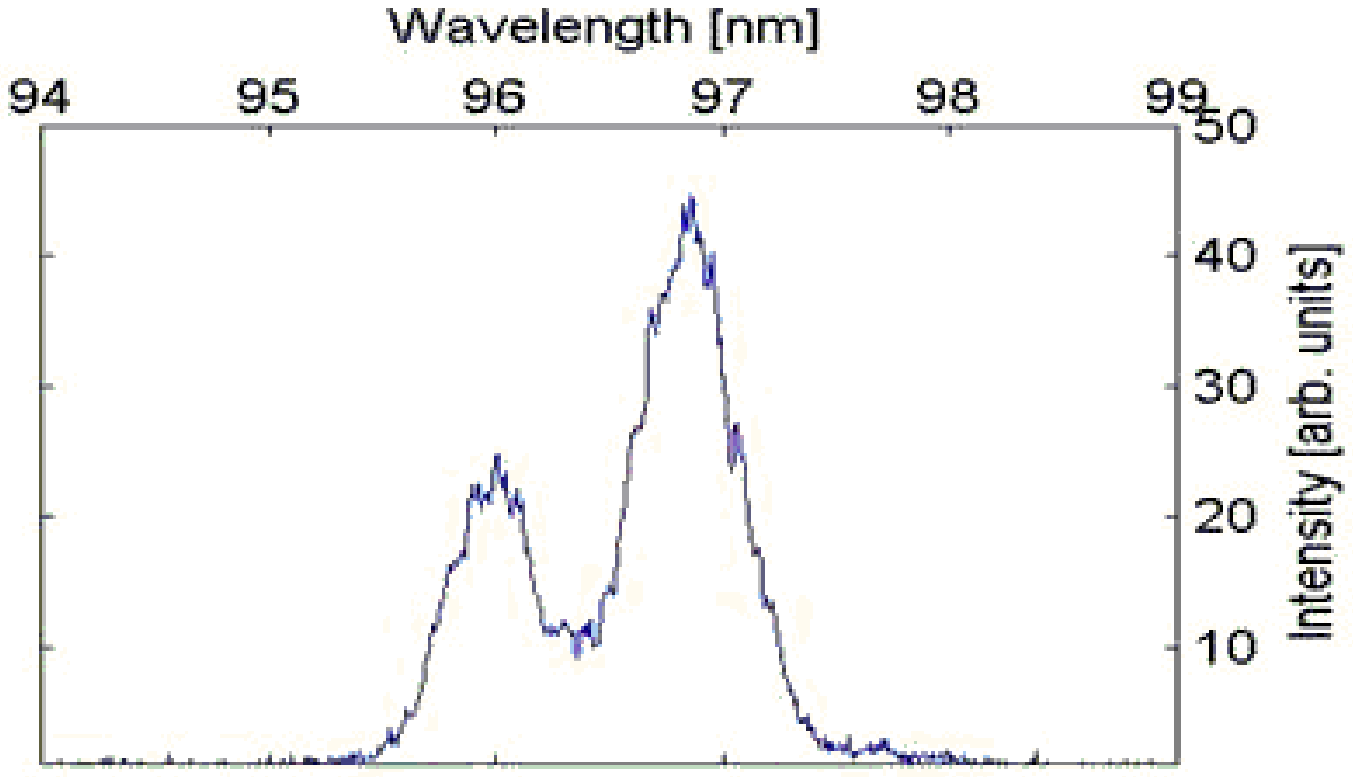}

\includegraphics[width=0.5\textwidth]{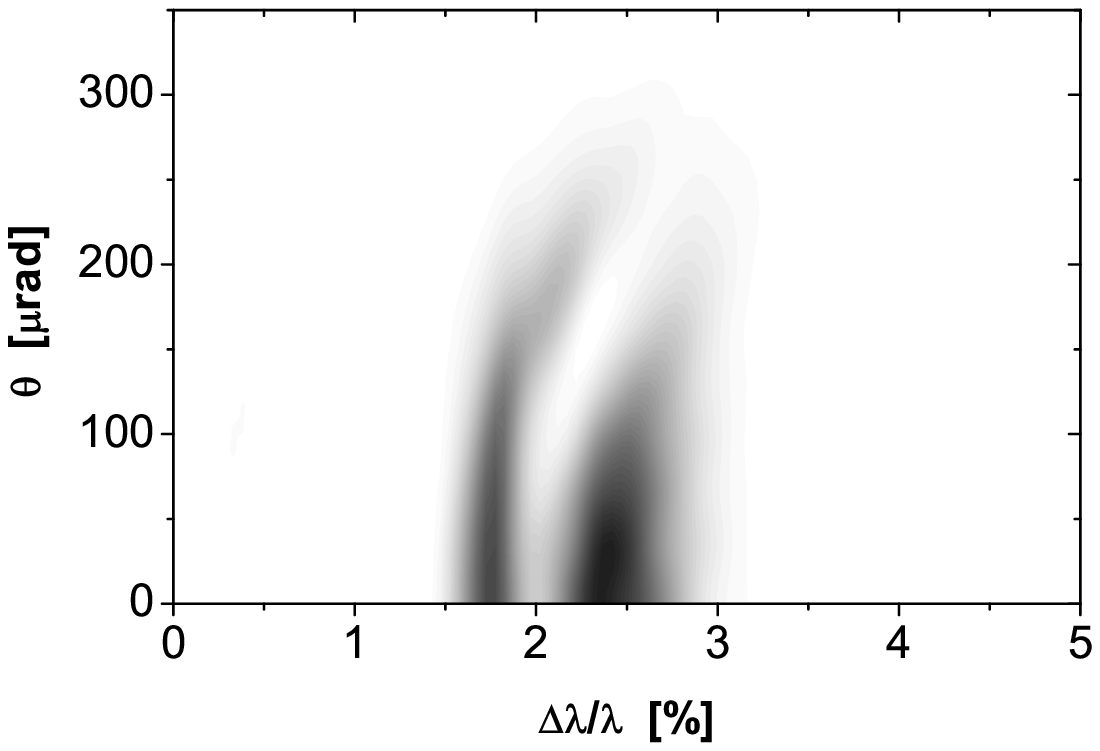}

\vspace*{-60mm}

\hspace*{0.5\textwidth}
\includegraphics[width=0.5\textwidth]{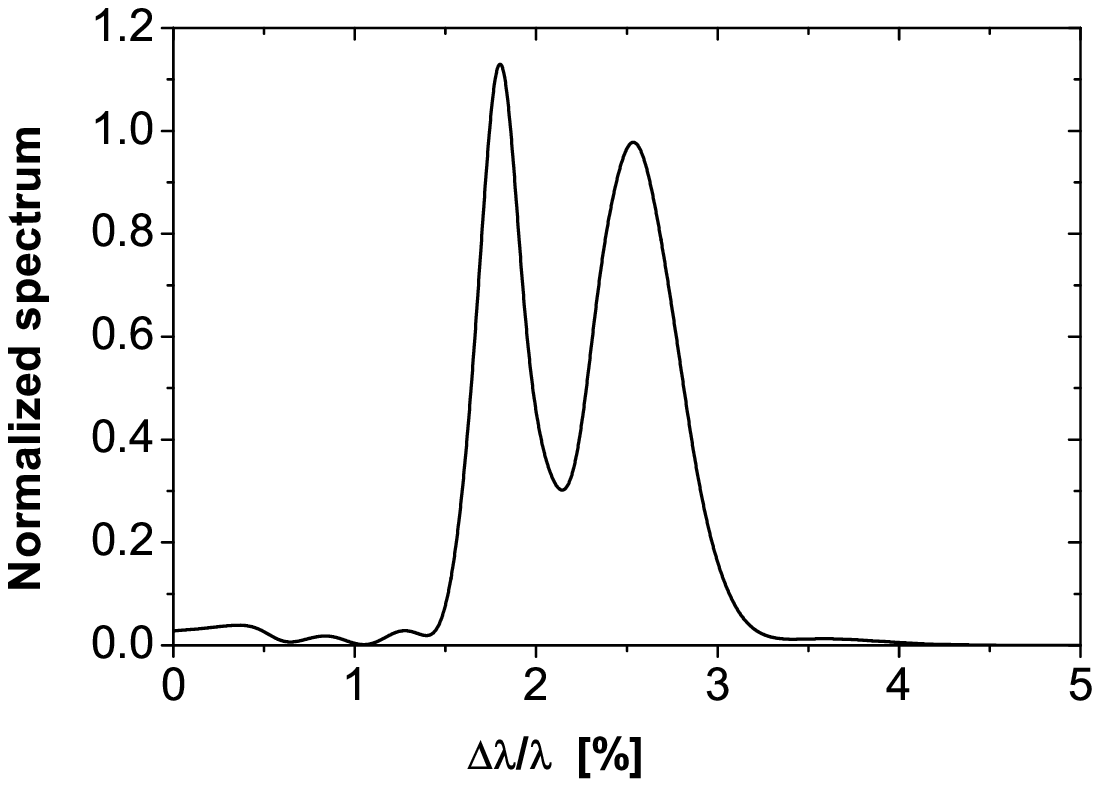}

\caption{\normalsize Single shot spectrum for TTF FEL operating in the nonlinear
regime. Left column shows image on CCD camera, and right column shows
projections.
Upper row are experimental data \cite{fel2002-stat} and lower row
represents results of simulations with code FAST
}
\label{fig:2sp}

\end{figure}

\subsection{Radiation spectra and fluctuations}

\begin{figure}[tb]

\includegraphics[width=0.5\textwidth]{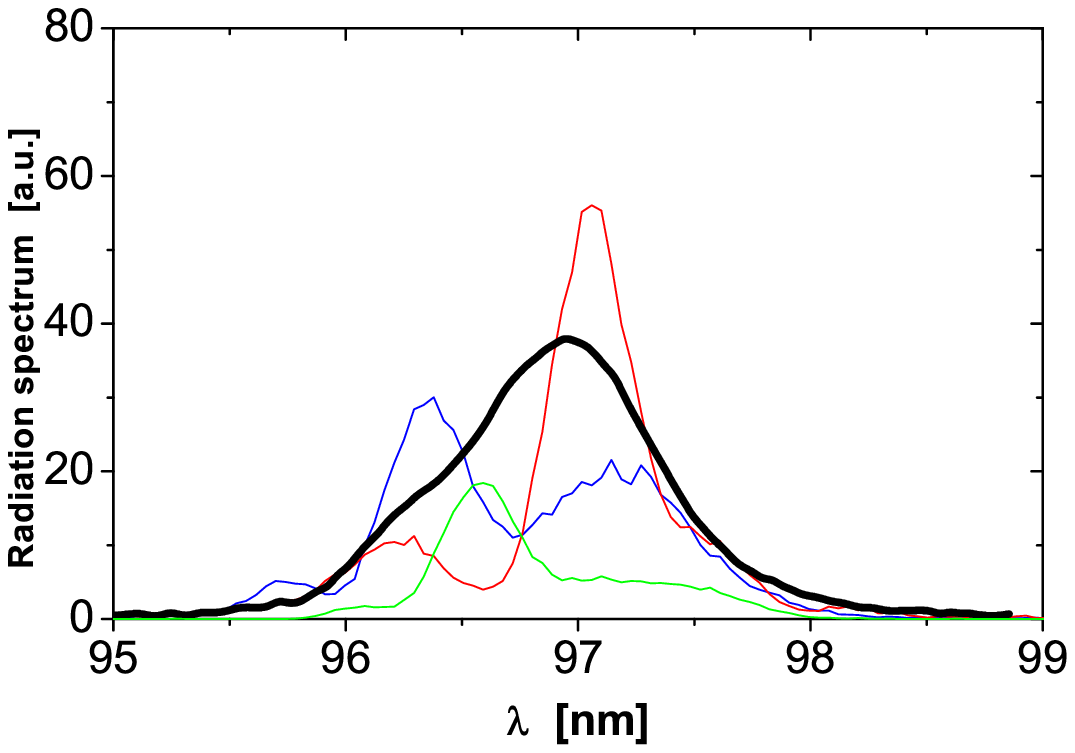}

\vspace*{-57mm}

\hspace*{0.5\textwidth}
\includegraphics[width=0.5\textwidth]{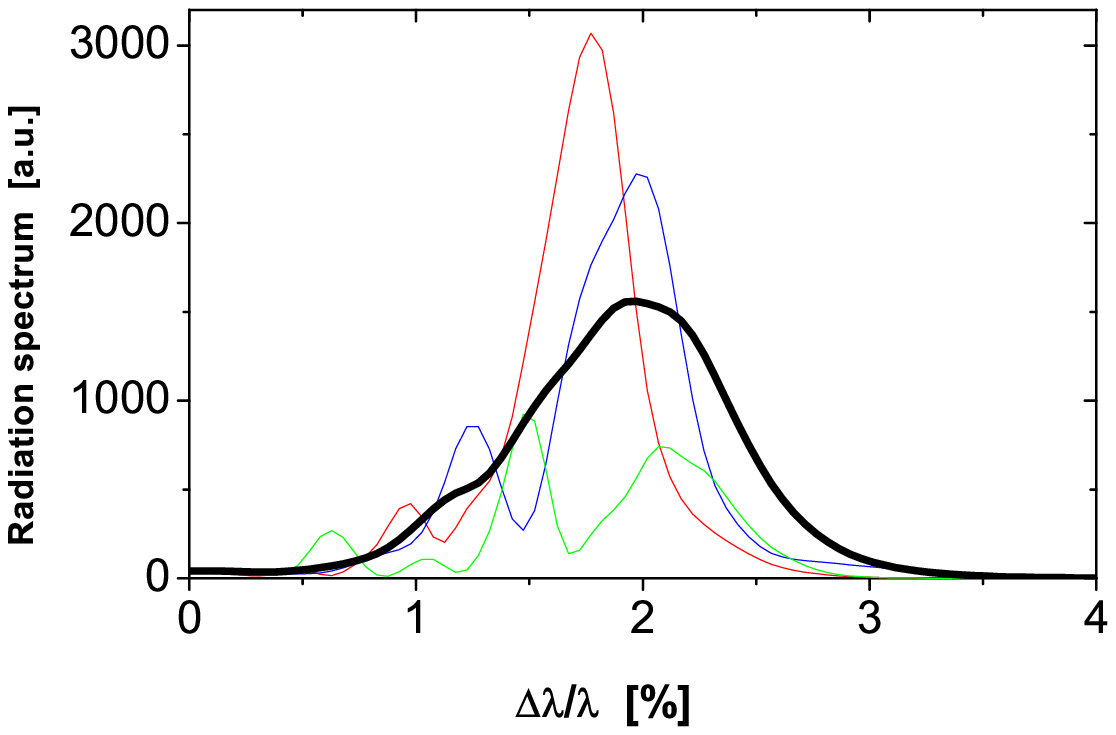}

\caption{\normalsize
Single shot spectra (thin lines) and averaged spectrum (bold lines) of
TTF FEL operating in the nonlinear regime.
Left plot shows experimental data \cite{ttf-sat-prl} and right plot
shows results of simulations with code FAST
}
\label{fig:sp-nonlin}

\includegraphics[width=0.5\textwidth]{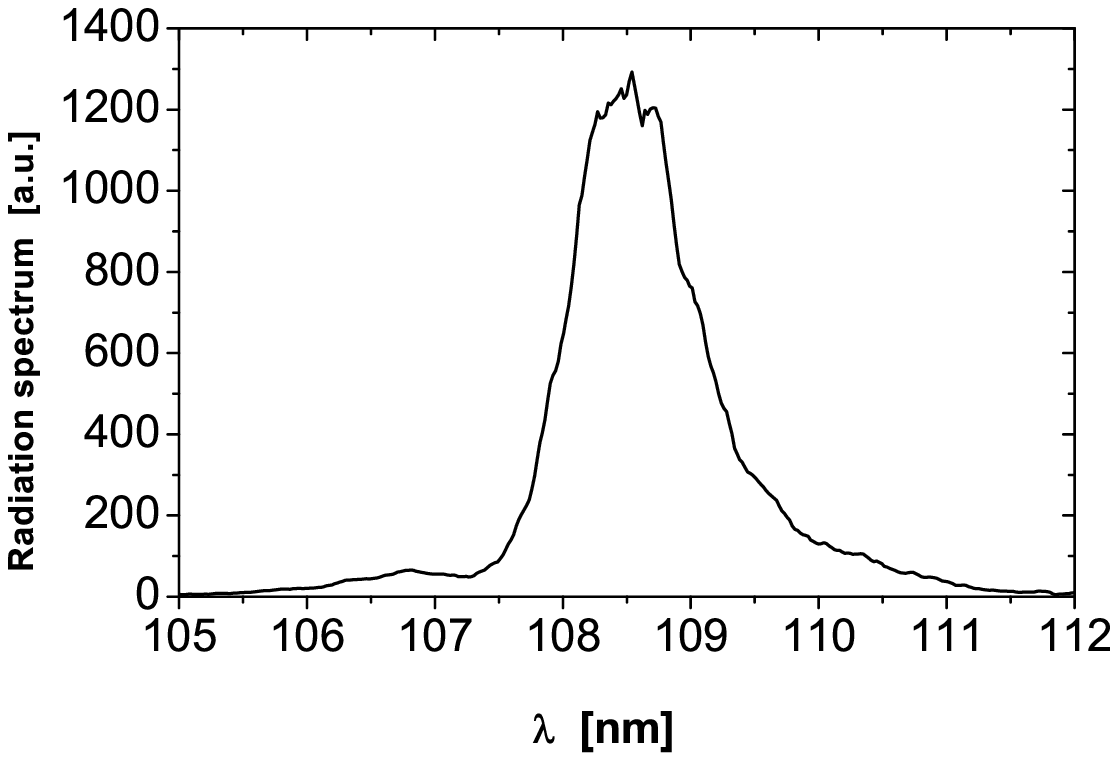}

\vspace*{-58mm}

\hspace*{0.5\textwidth}
\includegraphics[width=0.5\textwidth]{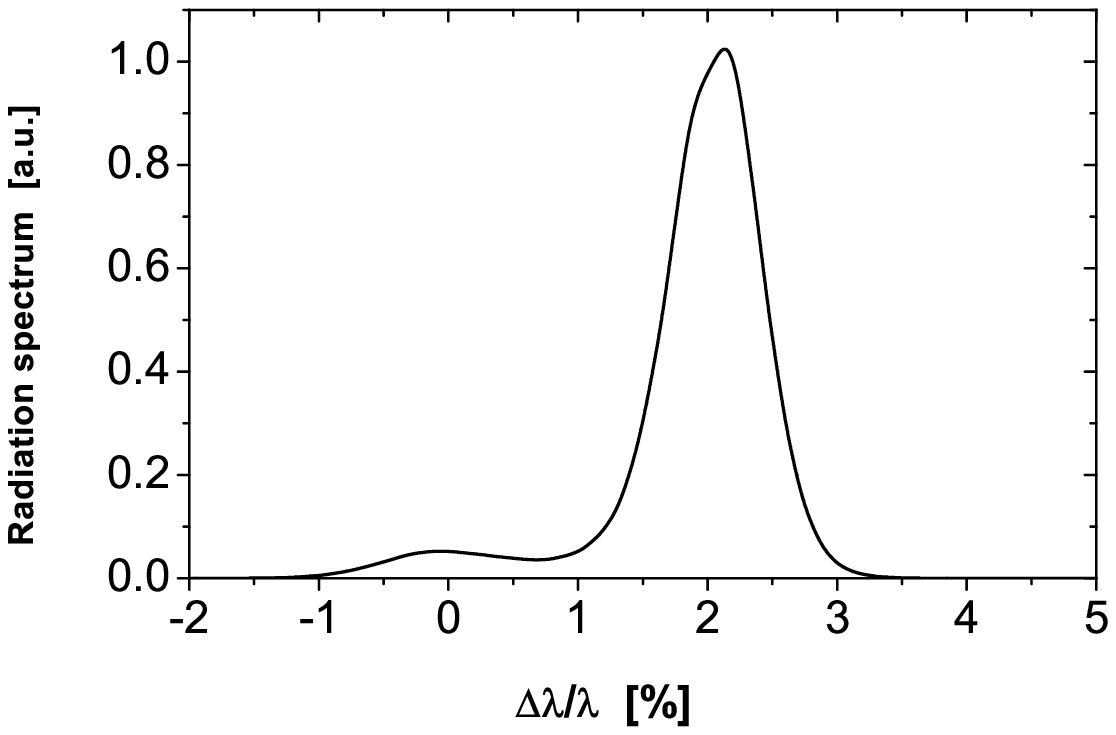}

\caption{\normalsize
Average spectrum of TTF FEL operating in the high-gain linear regime.
Left plot shows experimental data \cite{ttf-firstlasing-prl} and right
plot shows results of simulations with code FAST
}
\label{fig:spav-lin}

\end{figure}

The next topic of our study is the radiation spectra. Experimentally such
spectra can be measured in a single shot by the mean of a monochromator~\cite{monochr}.
In such a measurement the procedure is as follows: the photon beam is deflected by
plane mirror, passed through a narrow slit, and then  dispersed by a grating. The
dispersed beam is then detected by CCD camera as illustrated in  Fig.~\ref{fig:2sp}
in the left-corner plot. Projection of the CCD readings onto $x$-axis gives the spectrum
(see right-corner plot in Fig.~\ref{fig:2sp}). The same procedure was reproduced in the
simulations and gave the results presented in the lower plots of Fig.~\ref{fig:2sp} .
The range of spectral measurements was rather limited due to limited sensitivity of the
spectrometer. It was possible to detect reliably single shot spectra in
the nonlinear regime, while in the linear regime only spectra, averaged
over many shots are available after procedure of background
subtraction. In Figure~\ref{fig:sp-nonlin} we present comparison of experimental and
simulated spectra for nonlinear regime. We again see not only qualitative, but very
good quantitative agreement. Note that the bump on the left slope of the average spectra
was a signature for all spectral measurements at TTF FEL starting from the first lasing on
February 22, 2000 (see Fig.~\ref{fig:spav-lin}). Such a bump is indeed the consequence
of the strong energy chirp along the bunch.

\begin{figure}[p]
\begin{center}
\includegraphics[width=0.9\textwidth]{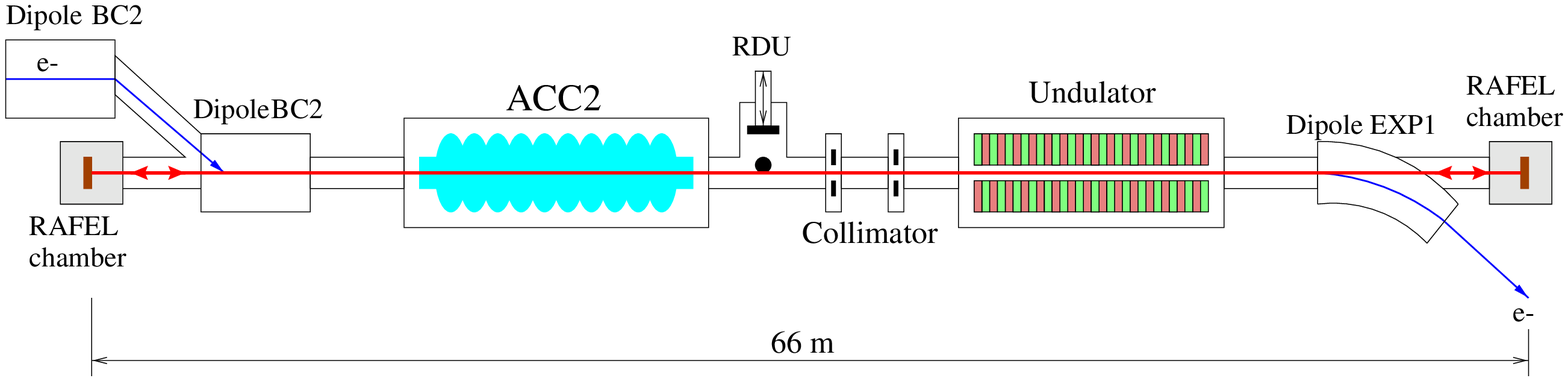}
\end{center}
\caption{
General layout of the experimental facility for measurement of
statistical properties of the radiation after narrow-band
monochromator. Here BC2 is the bunch compressor, ACC2 is the second
accelerating module, RDU denotes the MCP-based radiation
detector unit. RAFEL chamber downstream the undulator houses a plane
SiC mirror, and the chamber in the BC2 area hoses a grating.
}
\label{fig:layout-rafel}

\includegraphics[width=0.5\textwidth]{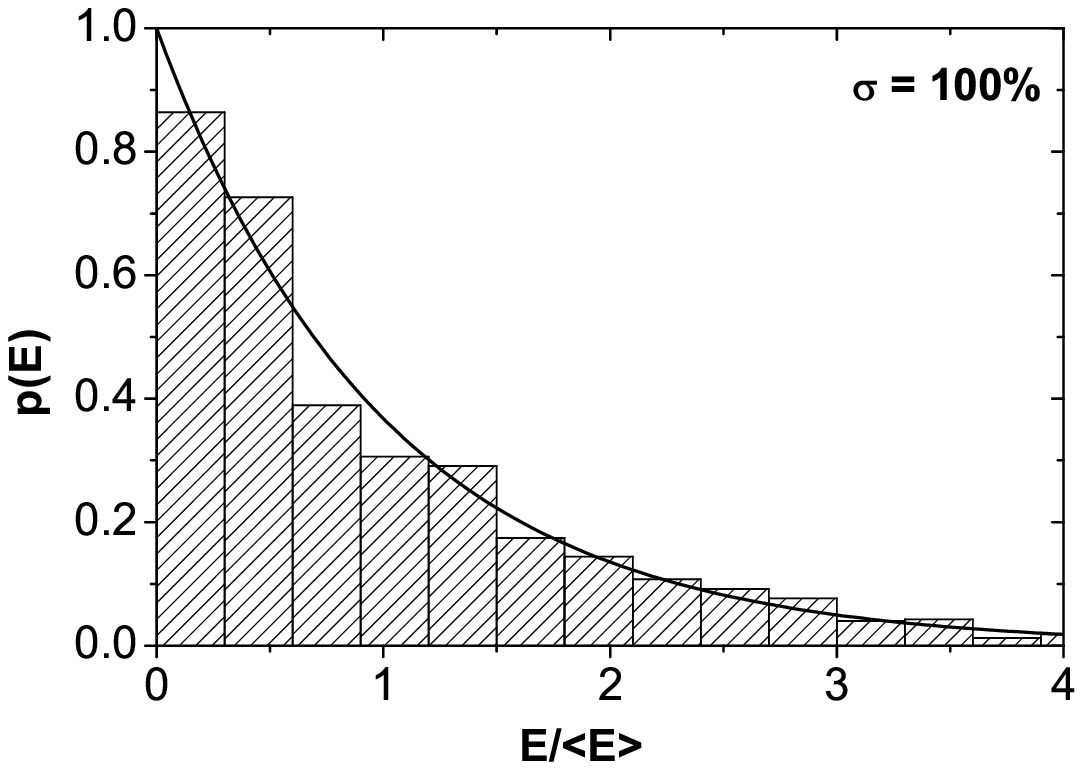}

\vspace*{-61mm}

\hspace*{0.5\textwidth}
\includegraphics[width=0.5\textwidth]{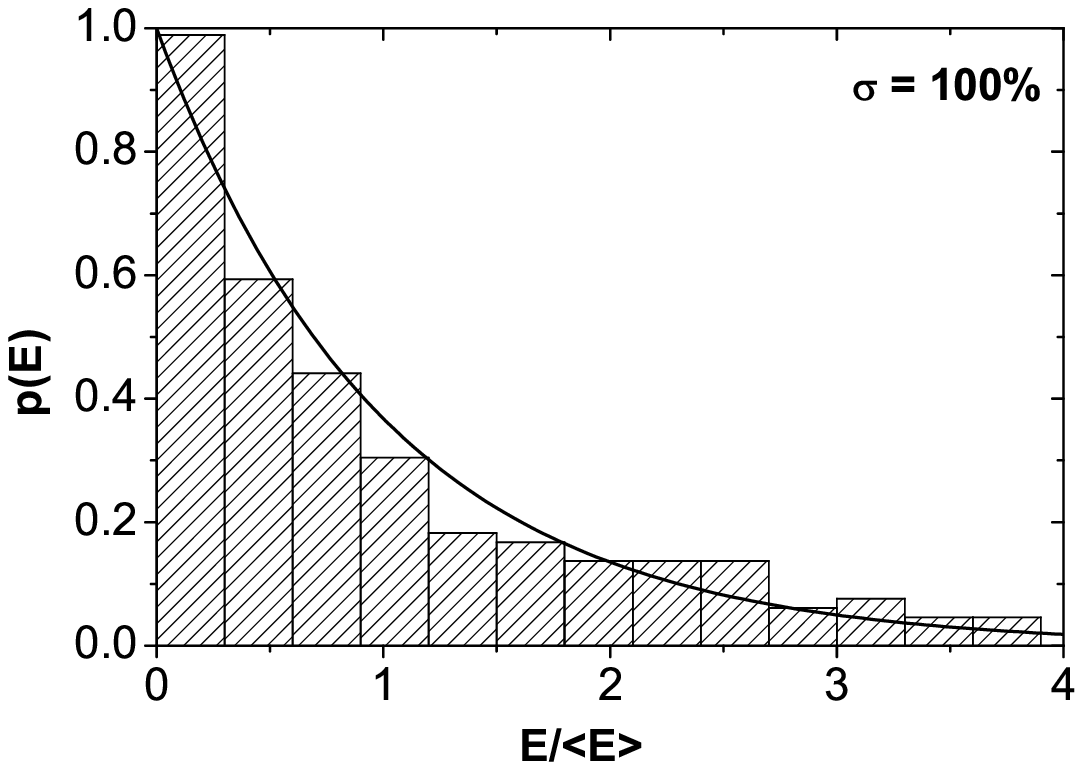}

\caption{\normalsize
Probability distribution of the energy in the radiation pulse after
narrow band monochromator.
TTF FEL operates in the linear regime.
Left plot shows experimental data with RAFEL grating
\cite{fel2002-stat} and right plot shows results of simulations with
code FAST.
Solid lines show negative exponential distribution
}
\label{fig:mon-1}

\includegraphics[width=0.5\textwidth]{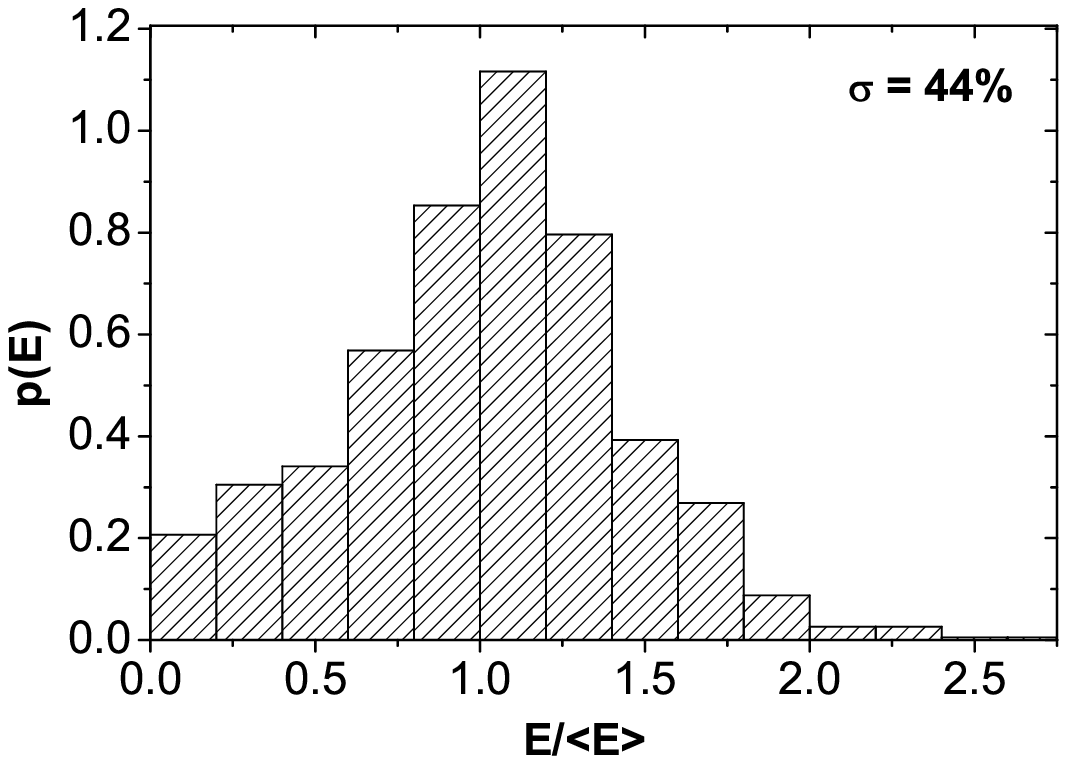}

\vspace*{-62mm}

\hspace*{0.5\textwidth}
\includegraphics[width=0.5\textwidth]{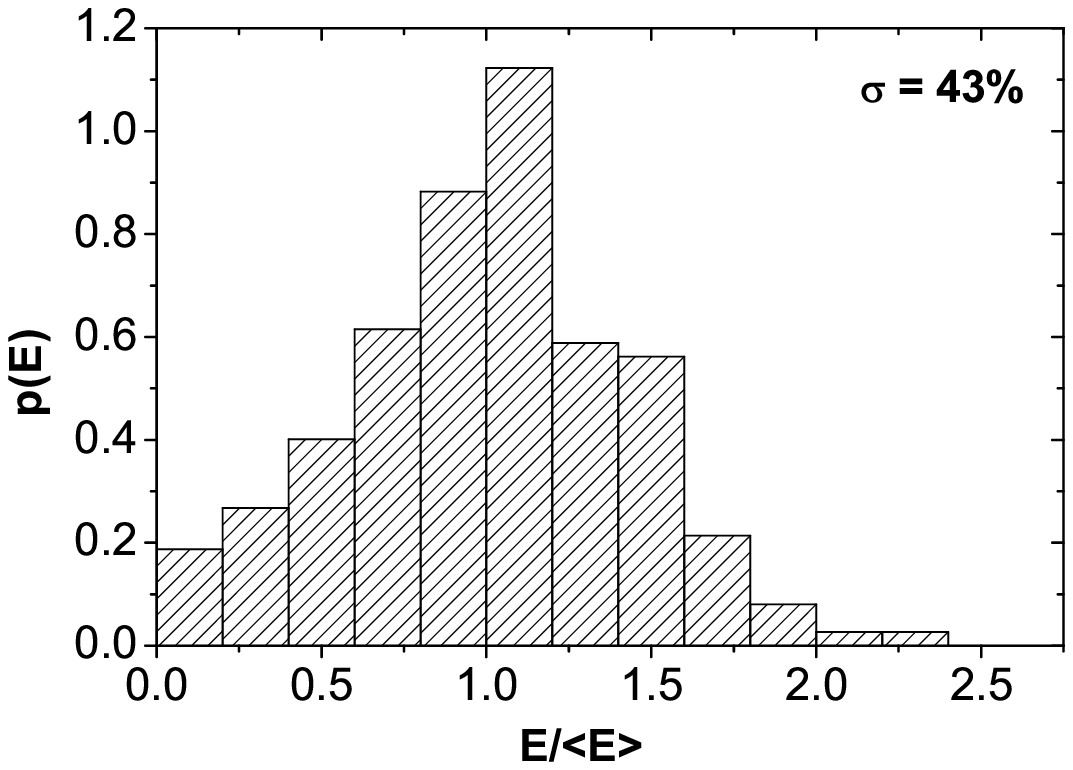}

\caption{\normalsize
Probability distribution of the energy in the radiation pulse after
narrow band monochromator.
TTF FEL operates in the nonlinear regime.
Left plot shows experimental data with RAFEL grating
\cite{fel2002-stat} and right plot shows results of simulations with
code FAST
}
\label{fig:mon-5}
\end{figure}

Another important topic of the FEL physics, well documented in the experiment, are
the statistical properties of the radiation after narrow-band monochromator.
Measurements of fluctuations of the radiation energy after the monochromator have
been performed using a narrow-band monochromator of
the RAFEL (Regenerative Amplifier FEL \cite{rafel-fel98}) optical
feedback system.  The scheme for these measurements is shown in Fig.~\ref{fig:layout-rafel}
\cite{fel2001-stat}. The SASE FEL radiation emitted by the electron beam is back-reflected
by a plane SiC mirror (RAFEL chamber at the right side of the scheme) onto
monochromator (RAFEL chamber at the left side of the scheme). The RAFEL
monochromator is a spherical grating in Littrow mounting which
disperses the light in the direction of the radiation detector unit
(RDU) installed 27~meters downstream. The RDU is equipped with an
MCP-based radiation detector with a thin (200~$\mu $m) gold wire which
plays the role of an exit slit of the monochromator. The design of a
spherical grating in Littrow mounting guarantees a resolution of about
$(\Delta \omega/\omega )_{\mathrm{M}} \simeq 10^{-4}$ which is much
less than typical scale of the spike in spectrum (see
Fig.~\ref{fig:sp-nonlin}). Wide dynamic range of MCP-based radiation
detector gave the possibility to perform reliable measurements for both the
linear and nonlinear regimes.

In the simulation procedure we also took into consideration aperture
limitations along the path of the photon beam
(9~mm in the undulator, and 6~mm in the spoiler of the electron beam
collimator). The reason for this is that diffraction effects at
aperture edges lead to mixing of spectrum which initially was strongly
correlated with the angle (see Fig.~\ref{fig:2sp}).

The plots in Fig.~\ref{fig:mon-1} show the probability distributions of the
radiation energy after a narrow-band monochromator. SASE FEL operates in the linear
regime, the active undulator length during this measurement was 9~m. When SASE FEL operates
in linear regime, the probability density must be a negative exponential distribution:

\begin{displaymath}
p(E) = \frac{1}
{\langle E \rangle }
\exp\left(-\frac{E}
{\langle E \rangle }\right) \ .
\end{displaymath}

\noindent The solid curves in Fig.~\ref{fig:mon-1} represent the negative
exponential distribution. So, we can conclude that both, measured
and simulated properties well follow the general properties typical for
completely chaotic polarized light.

Theory of SASE FEL predicts that in the saturation regime the
fluctuations should drop visibly when pulse duration $T$ is such that
$\rho \omega T \lesssim 2$ \cite{short-pulse}. At larger values of
$\rho \omega T $ fluctuations increase and quickly approach 100\%
level. Since radiation pulse length of TTF FEL is about two cooperation
length, we should expect significant suppression of fluctuations in the
nonlinear regime, down to 40\%.  It is seen in Fig.~\ref{fig:mon-5}
that measured fluctuations drop drastically in the nonlinear mode of
operation of TTF FEL. There is not only qualitative agreement, but also
quantitative agreement with calculated probability density distribution
function.  It is worth to mention that such a stabilization of
fluctuations is an independent indication for very short pulse duration
$T$ \cite{short-pulse}.

\section{Discussion}

\begin{figure}[tb]

\includegraphics[width=0.5\textwidth]{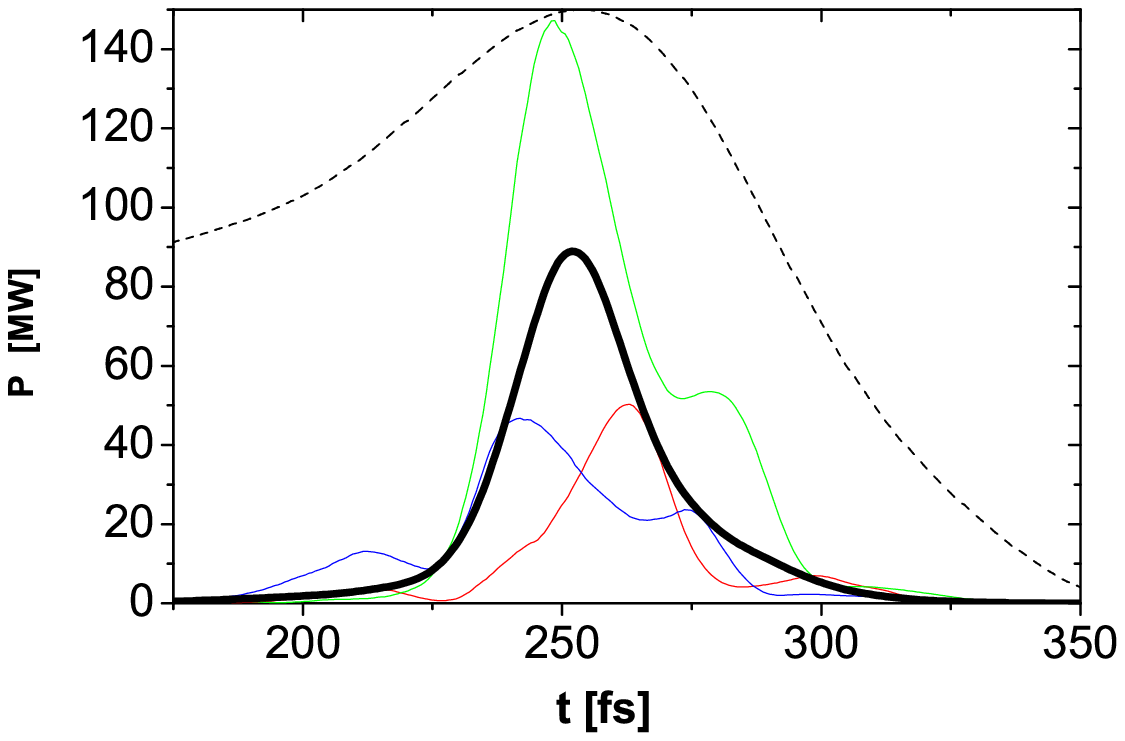}

\vspace*{-58mm}

\hspace*{0.5\textwidth}
\includegraphics[width=0.5\textwidth]{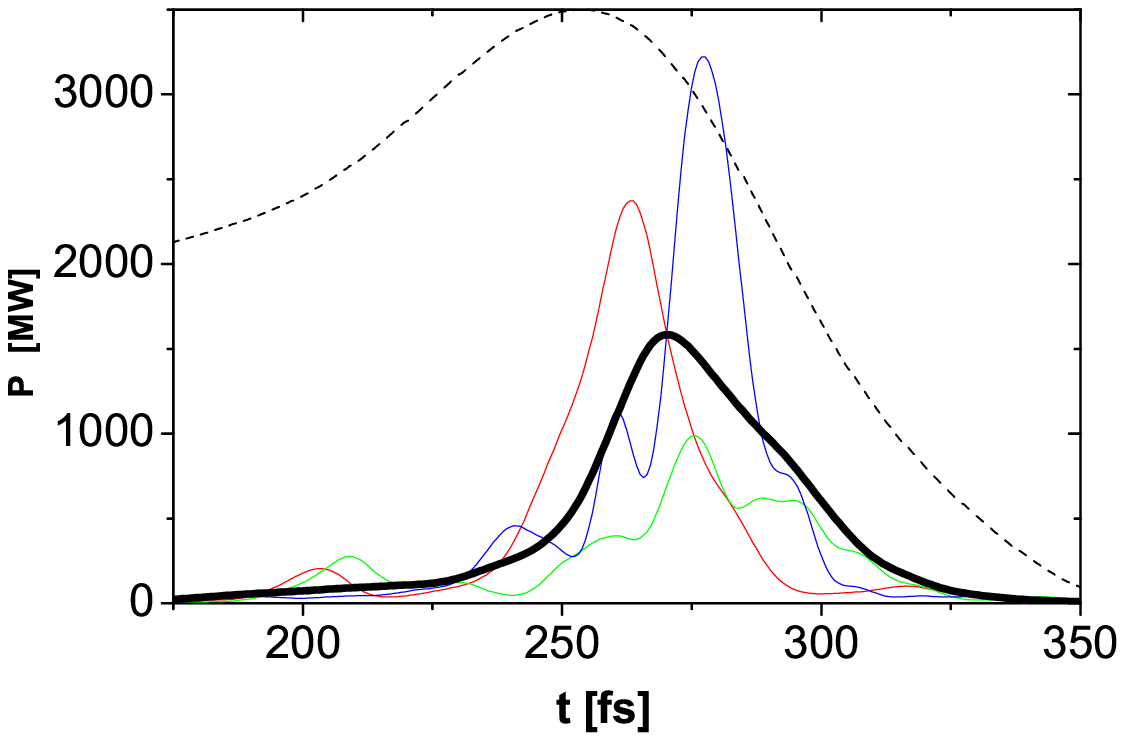}

\caption{\normalsize
Radiation power along the bunch for TTF FEL operating in the
linear (left plot) and nonlinear (right plot) regime.
Thin curves show single shots.
Bold curves show averaged profiles. Dashed curve shows
profile of electron bunch. Simulations are performed with
code FAST
 using bunch parameters shown in Fig.~\ref{fig:bunch-ue}
}
\label{fig:pav}
\end{figure}

The good agreement between experimental data and simulation results
allows us to determine the parameters of the FEL which are not directly
accessible experimentally. First of all this refers to the temporal
structure of the radiation pulse (see Fig.~\ref{fig:pav}): the computed
FWHM pulse duration in saturation regime is about 40~fs, and peak
power (averaged over ensemble) is 1.5  GW. We can also conclude that
the phase 1 of the SASE FEL at the TTF FEL, was driven by strongly
non-Gaussian beam having peak current about 3~kA. Beam dynamics in the
accelerator (even at high energy, after bunch compressor) was strongly
influenced by the space charge effects.

Our simulations show that rather poor transverse emittance of the
electron beam originated from the injector. With a large value of
emittance, the beam dynamics is not strongly influenced by CSR effects
in the bunch compressor. However CSR effects might be a crucial issue
for low-emittance beams as foreseen in future X-ray FELs. Strong
longitudinal space charge effect was not expected at TTF since local
energy spread before compression (parameter defining spike width and
peak current after compression) was strongly overestimated. In future
machines this effect can be avoided by an appropriate choice of
compression schemes. Therefore, the results of our simulations cannot
be directly scaled to more advanced accelerator designs for X-ray FELs.

In this paper we did a "font-end" comparison: we only compared the
results of the simulations with the radiation properties which have
been measured reliably and with high accuracy. Although direct
comparisons of the beam dynamics simulations with the measurements of
the slice parameters of the beam at different location along the
accelerator would have been more appropriate, there was impossible due
to the lack of adequate beam diagnostics. Our rich experimental
experience at TTF shows that tuning of the accelerator for the FEL
operation is very difficult task without information about relevant
beam parameters. Excellent properties of the FEL radiation were mainly
achieved by a global empirical optimization of the machine. Such a
procedure might be impossible in much larger and more complicated
accelerator systems for X-ray FELs with very large parameter space to
be tuned. This underlines the urgent necessity to develop electron beam
diagnostics tools which are crucial for the proper operation of X-ray
SASE FEL. In particular, a method of the peak current and bunch profile
reconstruction, based on detection of infrared coherent radiation from
an undulator, would perfectly fit this purpose \cite{desy-031}.
Reliable methods for measuring slice emittance and energy spread on a
femtosecond time scale should be developed, too.

\clearpage


\section*{Appendix: Parameters of the leading peak in density
distribution created in the bunch compressor due to RF nonlinearity}

An expression for the beam density distribution after compression,
including RF nonlinearity, was derived in \cite{li}. We present here the
simplified formulas for the parameters of the leading peak.

Let us consider a long bunch with the constant current $I_0$
accelerated off-crest in an RF accelerator with the RF wavelength
$\lambda$. An energy chirp along the beam is

\begin{displaymath}
E = E_0 \cos (\phi_0 + \Delta \phi)
\end{displaymath}

\noindent Here $\phi_0$ is the phase of a reference particle, and
$\Delta \phi = 2 \pi s/\lambda$. A position $s$ of a particle is
positive if it is moving in front of the reference particle. We assume
that $\Delta \phi \ll 1$ and expand the relative energy chirp up to the
second order:

\begin{equation}
\tilde{\delta} = \frac{E - E_0}{E_0} \simeq
- \Delta \phi \tan \phi_0 - \frac{(\Delta \phi)^2}{2}
\label{chirp}
\end{equation}

\noindent For a Gaussian uncorrelated energy spread a distribution
function in the vicinity of the reference particle can be described as
follows:

\begin{equation}
f (s,\delta) =
\frac{I_0}{\sqrt{2\pi} \sigma_{\delta}} \exp \left[ -
\frac{\left( \delta + 2 \pi s \tan(\phi_0) /\lambda +  2 \pi^2
s^2/\lambda^2 \right)^2}{2 \sigma_{\delta}^2} \right] \ ,
\label{distr-before}
\end{equation}

\noindent where $\sigma_{\delta}$ is the relative rms energy spread.
The normalization is chosen such
that after integration over $\delta$ we get the current.

Behind the bunch compressor a particle position $s_f$ (with respect to
a nominal particle) is connected with its position before compression
$s_i$ and energy deviation $\delta$ as

\begin{equation}
s_{f} \simeq s_{i} + R_{56} \delta + T_{566} \delta^2 + ... \ ,
\label{compr}
\end{equation}

\noindent where $R_{56}$ and $T_{566}$ are the first and the second
order momentum compaction. In our consideration we choose the
signs such that for a bunch compressor chicane $R_{56} > 0$ and
$T_{566} \simeq - 3 R_{56}/2$ (for a small bending angle). It is seen
from (\ref{chirp}) and (\ref{compr}) that the phase space distribution
after compression has a parabolic shape. The reference particle is
positioned at the fold-over when

\begin{displaymath}
\tan(\phi_0) = \frac{\lambda}{2 \pi R_{56}}
\end{displaymath}

\noindent Also, analyzing Eqs. (\ref{chirp}) and (\ref{compr}), we come
to the conclusion that the contribution of $T_{566}$ term can be
neglected under the condition $\phi_0 \ll 1$.
We also assume
that $\sigma_{\delta} \ll \phi_0^2$.
The final distribution function is obtained from (\ref{distr-before}) by a
simple substitution $s \to s - R_{56} \delta$ and is simplified with
the help of the above mentioned assumptions:

\begin{equation}
f (s,\delta) =
\frac{I_0}{\sqrt{2\pi} \sigma_{\delta}} \exp \left[ -
\frac{
\left( s/R_{56} +  2 \pi^2 R_{56}^2 \delta^2 /\lambda^2 \right)^2 }
{2 \sigma_{\delta}^2} \right] \ .
\label{distr-after}
\end{equation}

To obtain the dependence of current on s after compression one should
make the integration:

\begin{displaymath}
I(s) = \int \limits_{-\infty}^{\infty} \d \delta \ f(s,\delta)
\end{displaymath}

\noindent After normalization we get:

\begin{equation}
I(\hat{s}) = I_0 C g_1(\hat{s}) \ ,
\label{curr}
\end{equation}

\noindent where

\begin{displaymath}
\hat{s} = \frac{s}{R_{56} \sigma_{\delta}} \ , \ \ \ \
C = \frac{\lambda}{ \pi R_{56}
\sqrt{2\sigma_{\delta}}} \ ,
\end{displaymath}

\noindent and the function $g_1(\hat{s})$ is

\begin{equation}
g_1(\hat{s}) = \sqrt{\frac{2}{\pi}}
\int \limits_{0}^{\infty} \d x
\exp \left[ -
\frac{(\hat{s} + x^2)^2 }{2} \right] \ .
\label{g1}
\end{equation}

\begin{figure}[tb]
\begin{center}
\epsfig{file=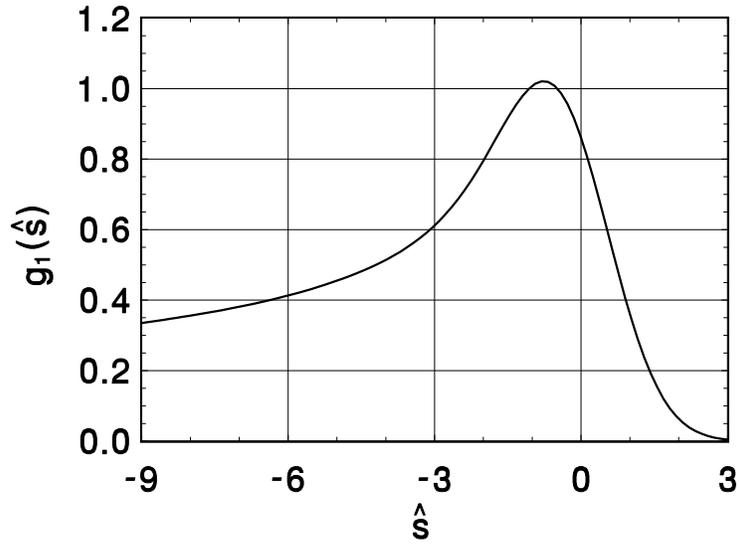,width=0.65\textwidth}
\end{center}
\caption{Function $g_1(\hat{s})$
}
\label{fig:g1}
\end{figure}

\noindent The plot of the function $g_1(\hat{s})$ is presented in Fig.~\ref{fig:g1}.
The maximal value of the function
$g_1(\hat{s})$ is close to 1, $\mathrm{max}(g_1(\hat{s})) \simeq 1.02$, so
that $C$ describes the enhancement of current. The full width (at half
maximum) of this curve is 4.8, or in dimensional notations:

\begin{equation}
(\Delta s)|_{\small \mathrm{FWHM}} \simeq 4.8 R_{56} \sigma_{\delta}
\label{fwhm}
\end{equation}

\noindent At the level of 0.8 the full width $\Delta \hat{s}$ is equal
to 2. We can also estimate the length of the beam slice
before compression, contributing to the leading peak after compression
- it is of the order of $\lambda \sqrt{\sigma_{\delta}}$. Thus, if the
current (before compression) only weakly
changes on this scale, our approximation of constant current is valid.

\noindent It is interesting to know the slice energy spread after
compression. It is easy to obtain from (\ref{distr-after}) the
following result:

\begin{equation}
\delta_{\mathrm{rms}}(\hat{s}) = \sigma_{\delta} C g_2(\hat{s}) \ ,
\label{en-spread}
\end{equation}

\noindent where

\begin{equation}
g_2(\hat{s}) = \left[ \frac{
\int \limits_{0}^{\infty} \d x x^2
\exp \left[ -
(\hat{s} + x^2)^2/2 \right] }
{ \int \limits_{0}^{\infty} \d x
\exp \left[ -
(\hat{s} + x^2)^2/2 \right] }
\right]^{1/2}
\label{g11}
\end{equation}

\begin{figure}[tb]
\begin{center}
\epsfig{file=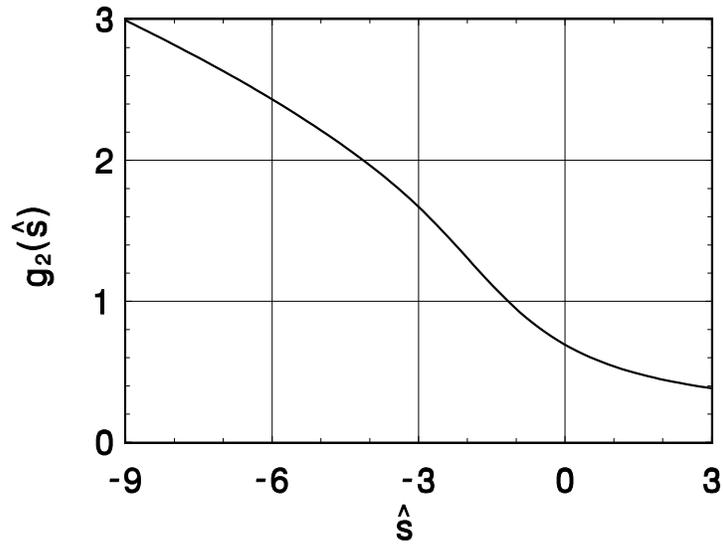,width=0.65\textwidth}
\end{center}
\caption{Function $g_2(\hat{s})$
}
\label{fig:g2}
\end{figure}

\begin{figure}[tb]
\begin{center}
\epsfig{file=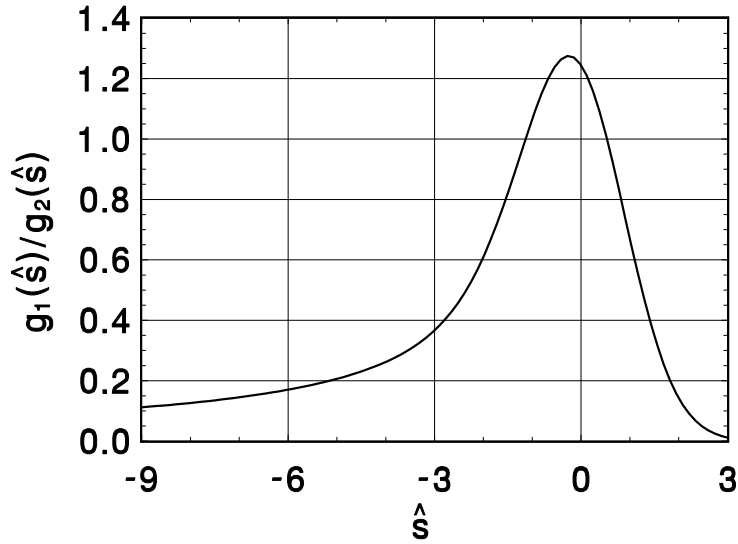,width=0.65\textwidth}
\end{center}
\caption{The ratio $g_1(\hat{s})/g_2(\hat{s})$
}
\label{fig:g12}
\end{figure}

\noindent The plot of this function is presented in Fig.~\ref{fig:g2} In the slice
with maximum current, $\hat{s} = -0.8$, the function $g_2$ takes the
value of 0.89. The ratio\footnote{We do not call this quantity
"longitudinal brightness" to avoid a possible confusion since the ratio
of local current to slice energy spread does not have to be conserved.
Instead, the phase space density is conserved. Note also that in our
simplified treatment we did not take care of the simplecticity of
transformation. Nevertheless, this practically does not influence our
results.}

\begin{displaymath}
\frac{I(s)}{\delta_{\mathrm{rms}}(s)} =
\frac{I_0}{\sigma_{\delta}} \frac{g_1(\hat{s})}{g_2(\hat{s})}
\end{displaymath}

\noindent can be important for FEL operation. In Fig.~\ref{fig:g12} we plot the
function $g_1(\hat{s})/g_2(\hat{s})$. One can see some
enhancement of the ratio of local current to slice energy spread with
respect to uncompressed beam. The full width of the curve is
about 3, beyond this good range it quickly goes down, so that only a
small part of the beam is favorable for lasing (if it is not spoiled
by collective effects). Note that rms energy spread is an adequate
quantity for SASE FEL calculation when it is much less than the FEL
parameter $\rho$ \cite{bonifacio,book}. When it becomes comparable to $\rho$,
one should take into account the actual distribution (to know only
dispersion is not sufficient). When it is much larger that $\rho$, one
can think of two independent beams, having different energies and
twice lower current.

Finally, let us present a numerical example. An electron beam with the
current 80 A and rms energy spread 3.5 keV is accelerated up to 135 MeV
in the RF accelerator with $\lambda = 23$ cm. After that it is
compressed in a bunch compressor with $R_{56} = 23$ cm. The peak
current after compression is then 3.6 kA, rms energy spread in
the slice with maximum peak current is about 140 keV, and full width of
the leading peak is 28 $\mu$m.

\end{document}